\pdfoutput=1

\documentclass[11pt,letterpaper,hyperref,dvipsnames]{cernyrep}
\usepackage[utf8]{inputenc}

\DeclareUnicodeCharacter{2212}{-}
\DeclareUnicodeCharacter{202F}{\,}

\hypersetup{
  colorlinks=true,
  linkcolor=red,
  urlcolor=blue,
  citecolor=darkgreen,
  pdfauthor={LH21 participants},
  pdftitle={Les Houches 2021 Proceedings}
}

\usepackage{cite}
\usepackage{amsmath,amssymb,graphicx,xspace,subfigure}
\usepackage{mathtools}
\usepackage{wasysym}
\usepackage{bigints}
\usepackage{wrapfig}

\usepackage[T1]{fontenc}
\usepackage{lmodern}
\usepackage{rotating}
\usepackage{xcolor}
\usepackage{booktabs}
\usepackage{arydshln}
\usepackage{lscape}
\usepackage{floatpag}
\floatpagestyle{plain}
\usepackage{lineno}
\usepackage{graphpap}

\usepackage{import}
\usepackage{enumitem}
\usepackage{caption}
\usepackage{xspace}
\usepackage{adjustbox}
\usepackage{tikz}
\usetikzlibrary{positioning,arrows}
\usetikzlibrary{decorations.pathmorphing}
\usetikzlibrary{decorations.markings}

\newcommand{\eV}{\ensuremath{\text{e\kern-0.15ex{}V}}\xspace}

\newcommand{\TeV}{\ensuremath{\text{T\eV}}\xspace}

\definecolor{darkred}{rgb}{.8, 0.1, 0.1}
\definecolor{darkyellow}{rgb}{0.45, 0.45, 0.}
\definecolor{violet}{rgb}{1.0, 0.0, 1.0}
\definecolor{darkgreen}{rgb}{0.15, .8, 0.15}

\makeatletter
\renewcommand{\thechapter}{\@Roman\c@chapter}
\makeatother

\begin{document}

{\centering{\LARGE{\bf{Les Houches 2021 - Physics at TeV Colliders:\\ Report on the Standard Model Precision Wishlist \par }}}}

\pagenumbering{roman}

\vspace{0.7cm}

Alexander~Huss$^{1\,}$, Joey~Huston$^{2\,}$, Stephen~Jones$^{3\,}$, Mathieu~Pellen$^{4\,}$

\vspace{0.7cm}

\noindent{\small\it $^1$Theoretical Physics Department, CERN,} \\ %
{\small\it 1211 Geneva 23, Switzerland}\\[3mm]
\noindent{\small\it $^2$Department of Physics and Astronomy, Michigan State University,} \\ %
{\small\it East Lansing, MI 48824, USA}\\[3mm]
\noindent{\small\it $^3$Institute for Particle Physics Phenomenology, Durham University,} \\ %
{\small\it Durham DH1 3LE, United Kingdom}\\[3mm]
\noindent{\small\it $^4$Albert-Ludwigs-Universit\"at Freiburg, Physikalisches Institut,} \\ %
{\small\it Hermann-Herder-Stra\ss e 3, D-79104 Freiburg, Germany}\\[3mm]

{\leftline{\bf{Abstract}}}

\vspace{0.5cm}

Les Houches activities in 2021 were truncated due to the lack of an in-person component. However, given the rapid progress in the field and the restart of the LHC, we wanted to continue the bi-yearly tradition of updating the standard model precision wishlist. 
In this work we therefore review recent progress (since Les Houches 2019) in fixed-order computations for LHC applications.
In addition, necessary ingredients for such calculations such as parton distribution functions, amplitudes, and subtraction methods are discussed.
Finally, we indicate processes and missing higher-order corrections that are required to reach the theoretical accuracy that matches the anticipated experimental precision.

\newpage
\pagenumbering{arabic}
\setcounter{footnote}{0}


\newpage



\newcommand{\NLLgen}[1]{N${}^{#1}$LL\xspace}

\newcommand{\NLL}[1]{N${}^{#1}$LL\xspace}
\newcommand{\NLLone}{NLL\xspace}

\newcommand{\LL}{LL\xspace}
\newcommand{\NNLL}{NNLL\xspace}
\newcommand{\NNLLp}{NNLL'\xspace}
\newcommand{\NNNLL}{\NLLgen3\xspace}
\newcommand{\NNNLLp}{\NLLgen3'\xspace}

\newcommand{\NLO}[1]{N${}^{#1}$LO\xspace}
\newcommand{\NLOone}{NLO\xspace}
\newcommand{\NLOgen}{NLO\xspace}
\newcommand{\NNLOgen}{NNLO\xspace}
\newcommand{\NNNLOgen}{\NLO3}

\newcommand{\NLOH}[1]{N${}^{#1}$LO${}_{\rm HTL}$\xspace}
\newcommand{\NLOHone}{NLO${}_{\rm HTL}$\xspace}
\newcommand{\NLOHTL}{NLO${}_{\rm HTL}$\xspace}
\newcommand{\NNLOHTL}{NNLO${}_{\rm HTL}$\xspace}
\newcommand{\NNNLOHTL}{\NLOH3}

\newcommand{\NLOQ}[1]{N${}^{#1}$LO${}_{\rm QCD}$\xspace}
\newcommand{\LOQ}{LO${}_{\rm QCD}$\xspace}
\newcommand{\NLOQone}{NLO${}_{\rm QCD}$\xspace}
\newcommand{\LOQCD}{LO${}_{\rm QCD}$\xspace}
\newcommand{\NLOQCD}{NLO${}_{\rm QCD}$\xspace}
\newcommand{\NNLOQCD}{NNLO${}_{\rm QCD}$\xspace}
\newcommand{\NNNLOQCD}{\NLOQ3}

\newcommand{\NLOE}[1]{N${}^{#1}$LO${}_{\rm EW}$\xspace}
\newcommand{\NLOEone}{NLO${}_{\rm EW}$\xspace}
\newcommand{\LOEW}{LO${}_{\rm EW}$\xspace}
\newcommand{\NLOEW}{NLO${}_{\rm EW}$\xspace}
\newcommand{\NNLOEW}{NNLO${}_{\rm EW}$\xspace}

\newcommand{\NLOD}[1]{N${}^{#1}$LO${}_{\rm QED}$\xspace}
\newcommand{\NLODone}{NLO${}_{\rm QED}$\xspace}
\newcommand{\NLOQED}{NLO${}_{\rm QED}$\xspace}
\newcommand{\NNLOQED}{NNLO${}_{\rm QED}$\xspace}

\newcommand{\NLOSM}{NLO${}_{\rm SM}$\xspace}

\newcommand{\NLOQE}[2]{N${}^{(#1,#2)}$LO${}_{{\rm QCD}\otimes{\rm EW}}$\xspace}

\newcommand{\NLOHE}[2]{N${}^{(#1,#2)}$LO${}^{\rm (HTL)}_{{\rm QCD}\otimes{\rm EW}}$\xspace}

\newcommand{\NLOmixQED}[2]{N${}^{(#1,#2)}$LO${}_{{\rm QCD}\otimes{\rm QED}}$\xspace}

\newcommand{\NLOQmtsix}[1]{N${}^{#1}$LO${}_{\rm QCD}^{(1/{m_t^8})}$\xspace}
\newcommand{\NLOQzzero}[1]{N${}^{#1}$LO${}_{\rm QCD}^{(z\to0)}$\xspace}
\newcommand{\NLOQVBF}[1]{N${}^{#1}$LO${}_{\rm QCD}^{(\rm VBF)}$\xspace}
\newcommand{\NLOQoneVBF}{NLO${}_{\rm QCD}^{(\rm VBF)}$\xspace}
\newcommand{\NLOQCDVBF}{NLO${}_{\rm QCD}^{(\rm VBF)}$\xspace}
\newcommand{\NNLOQCDVBF}{NNLO${}_{\rm QCD}^{(\rm VBF)}$\xspace}
\newcommand{\NLOQoneDIS}{NLO${}_{\rm QCD}^{(\rm DIS)}$\xspace}
\newcommand{\NLOQDIS}[1]{N${}^{#1}$LO${}_{\rm QCD}^{(\rm DIS)}$\xspace}
\newcommand{\NLOEoneVBF}{NLO${}_{\rm EW}^{(\rm VBF)}$\xspace}
\newcommand{\NLOEWVBF}{NLO${}_{\rm EW}^{(\rm VBF)}$\xspace}

\newcommand{\NLOQoneVBFstar}{NLO${}_{\rm QCD}^{(\rm VBF^{*})}$\xspace}
\newcommand{\NLOQVBFstar}[1]{N${}^{#1}$LO${}_{\rm QCD}^{(\rm VBF^{*})}$\xspace}
\newcommand{\NLOQCDVBFstar}{NLO${}_{\rm QCD}^{(\rm VBF^{*})}$\xspace}
\newcommand{\NNLOQCDVBFstar}{NNLO${}_{\rm QCD}^{(\rm VBF^{*})}$\xspace}
\newcommand{\NNNLOQCDVBFstar}{\NLOQVBFstar3}

\newcommand{\NLOEoneVBFstar}{NLO${}_{\rm EW}^{(\rm VBF^{*})}$\xspace}
\newcommand{\NLOggHVtb}[1]{N${}^{#1}$LO${}_{gg\to HZ}^{(t,b)}$\xspace}
\newcommand{\NNLOQCDT}{NNLO${}_{\rm QCD}^{(t)}$\xspace}
\newcommand{\NNLOQCDBC}{NNLO${}_{\rm QCD}^{(b,c)}$\xspace}

\newcommand{\xs}{$\sigma$}
\newcommand{\tb}{\bar{t}}
\newcommand{\bb}{\bar{b}}
\newcommand{\qb}{\bar{q}}

\newcommand{\wodecay}{(w/o decay)}
\newcommand{\wdecay}{}
\newcommand{\wodecays}{(w/o decays)}
\newcommand{\wdecays}{}
\newcommand{\wleptdecays}{}

\newcommand{\MadgraphaMCatNLO}{\textsc{Madgraph5}\_a\textsc{MC@NLO}\xspace}
\newcommand{\OpenLoops}{O\protect\scalebox{0.8}{PENLOOPS}\xspace}
\newcommand{\Recola}{R\protect\scalebox{0.8}{ECOLA}\xspace}
\newcommand{\GoSam}{G\protect\scalebox{0.8}{O}S\protect\scalebox{0.8}{AM}\xspace}
\newcommand{\MadLoop}{M\protect\scalebox{0.8}{AD}L\protect\scalebox{0.8}{OOP}\xspace}
\newcommand{\Powheg}{P\protect\scalebox{0.8}{OWHEG}\xspace}
\newcommand{\Powhegboxres}{P\protect\scalebox{0.8}{OWHEG-BOX-RES}\xspace}
\newcommand{\PowhegboxVtwo}{P\protect\scalebox{0.8}{OWHEG-BOX-V2}\xspace}
\newcommand{\Herwig}{H\protect\scalebox{0.8}{ERWIG}\xspace}
\newcommand{\Matrix}{M\protect\scalebox{0.8}{ATRIX}\xspace}
\newcommand{\Munich}{M\protect\scalebox{0.8}{UNICH}\xspace}
\newcommand{\Geneva}{G\protect\scalebox{0.8}{ENEVA}\xspace}
\newcommand{\Sherpa}{S\protect\scalebox{0.8}{HERPA}\xspace}
\newcommand{\NNLOjet}{NNLO\protect\scalebox{0.8}{JET}\xspace}
\newcommand{\MiNLO}{M\protect\scalebox{0.8}{iNLO}\xspace}
\newcommand{\NLOX}{NLOX\xspace}

\section{Introduction}
\label{sec:SM_wishlist}

Identifying key observables and processes that require improved theoretical input has been
a key part of the Les Houches programme. In this contribution we briefly summarise progress since the previous
report in 2019 and explore the possibilities for further advancements.
We also provide an estimate of the experimental uncertainties for a few key processes. 
A summary of this sort is perhaps unique in the field and serves a useful purpose for both practitioners in the field and for other interested readers. Given the amount of work that has been, and is being, done, this summary will no doubt be incomplete, and we apologize for any omissions.%
\footnote{The Les Houches Disclaimer.}
Even though Les Houches 2021 did not take place due to the COVID pandemic, we felt that the tradition of updating the wishlist, and of summarizing the advances in LHC-related calculations, should continue. The Les Houches wishlist has served as a summary/repository/grab-bag for the state-of-the-art for the higher-order QCD and EW calculations relevant for the LHC, and thus forms a useful resource for both LHC theorists and experimentalists. If nothing else, this reduces the level of effort that will be required for an update in the 2023 Les Houches report. 

The present report is organised as follows: in section~\ref{sec:HOT}, several ingredients needed for the calculation of higher orders are discussed.
They range from parton distribution functions to amplitude techniques and subtraction methods.
In the second section~\ref{sec:SM_wishlist:precision_wish_list}, an exhaustive review of the results that appeared since Les Houches 2019 is given with an accordingly updated wishlist.

\section{Higher-order techniques}
\label{sec:HOT}

While the years before the Les Houches 2019 report~\cite{Amoroso:2020lgh} had been marked
by significant progress in the production of \NNLOgen results in an almost industrial manner
with most useful $2\to2$ processes having been calculated,
the last two years have seen a saturation due to the unavailability of 2-loop amplitudes
beyond $2\rightarrow2$ scattering.
However, remarkable progress was achieved in this direction by several groups and approaches
culminating in the first $2\rightarrow3$ calculations of a hadron collider process.
Closely related is the huge progress in the calculation of 2-loop 5-point amplitudes,
as well as 2-loop amplitudes for $2\to2$ processes involving internal masses.
For a review of some recent developments see also Ref.~\cite{Heinrich:2020ybq}.

However, it is not only the amplitude community that has seen impressive development. There have also
been significant steps forward on the side of subtraction schemes, and there are in the meanwhile several subtraction and slicing methods available to deal (in principle) with higher-multiplicity processes at \NNLOgen (see below).

On the parton shower side, NLO QCD matched results and matrix element improved multi-jet merging techniques have become a standard level of theoretical precision.
The automation of full SM corrections including NLO electroweak predictions has also seen major improvements.

Another challenge is to make the \NNLOgen $2\to2$ predictions or complex NLO predictions publicly available to experimental analyses, and there has been major progress to achieve this goal. {\sc Root nTuples} have been a useful tool for complicated final states at NLO and
allow for very flexible re-weighting and analysis. The cost for this is
the large disk space required to store the event information.

Finally, an extension of APPLgrid~\cite{Carli:2010rw} and fastNLO~\cite{Kluge:2006xs} to NNLO~\cite{Britzger:2019kkb} offers a convenient method to distribute higher-order predictions.
Despite progress in this direction~\cite{Carrazza:2020gss}, there have been only a limited number of these grids made publicly available.
Once it is the case, they will likely be used heavily in precision PDF fits as well.

Below, we discuss some aspects of higher-order computations.

\subsection{Parton distribution functions}

One of the elements in improving the accuracy of theoretical predictions at the LHC lies in the determination of parton distribution functions (PDFs). PDFs are most commonly determined by global fits to experimental data, taking into account the experimental errors in the data. The standard now is for the PDFs to be determined at NNLO QCD, although fits at NLO QCD and LO are still available. It is encouraged to use NLO QCD (or even NNLO QCD) PDFs where possible, even for computation of lower perturbative accuracy. The results of the global fits are central values for each flavour PDF, along with an estimate of the PDF uncertainty, dominated by the input experimental errors for the data included in the fit. The formalism used in the fit can either be Hessian~\cite{Hou:2019efy,Bailey:2020ooq} or based on Monte Carlo replicas~\cite{NNPDF:2021njg}. The number of data points included in the global PDF fits is typically of the order of 3000--4000 from a wide range of processes.

Recently, many differential cross section measurements from the LHC have been included in the PDF determination.
This was made possible by the \NNLOQCD calculations of the relevant $2\rightarrow2$ matrix elements that have been discussed in past iterations of the wishlist.
For use in calculations at \NNNLOgen, several of which are discussed here, nominally \NNNLOgen PDFs would be needed. As they are not yet available, NNLO PDFs are used in their stead with an unknown uncertainty introduced into the predictions as a result. This has a non-negligible impact on the Higgs cross section at \NNNLOgen through gluon-gluon fusion, for example.  There are efforts to estimate the theoretical uncertainties due to (missing) higher order terms. These would be in addition to the (dominant) experimental uncertainties from the data included in the PDF fits. The theory uncertainties would be obtained by variations of the renormalization and factorization scales that are used to evaluate the matrix elements at \NNLOgen. Considering separate scales of each type for each data-set calculation would add too many degrees of freedom and remove much of the constraining power of the PDF fit. Connecting the renormalization or factorization scales, even for similar processes, may be treating those scales as more physical than they deserve. Perhaps there is more justification for treating the factorization scale in this manner than the renormalization scale. There is also the issue of whether introducing additional uncertainties in the PDFs through scale variations, and then in addition, performing scale variations in the predictions in the nominal manner, may lead to an over-counting of the uncertainty. Ref.~\cite{Harland-Lang:2018bxd} proposes using a physical basis (for example structure functions or similar observables) rather than the PDFs themselves. Considering correlated factorization scale variations in the PDF fit, and not in the resultant predictions, may not be ideal but an acceptable solution for certain specific physical quantities.

Ref.~\cite{NNPDF:2019ubu} proposes taking into account the missing higher order uncertainties in the cross sections included in the PDF fits by adding a theory uncertainty to the experimental covariance matrix. Since the theory uncertainties are uncorrelated with the experimental ones, the two uncertainties can be added in quadrature in the covariance matrix. The global fit processes are divided into five separate types (DIS NC, DIS CC, Drell-Yan, jets and top), with a hypothesis that calculations within a given type will be likely to
have similar structures of higher-order corrections. An assumption is made that the  renormalization scale
is only correlated within a single type of process, while the factorization scale is fully correlated across all processes. Resultant fits to the NNPDF3.1 data set (with some extra kinematic restrictions) do not substantially change the PDF uncertainties, but may relax some of the tensions among the experimental data sets.

MSHT
has carried out an exercise of parametrising the higher order effects with nuisance parameters based on a prior probability distribution (using the information currently available regarding \NNNLOgen matrix elements and the approximate splitting functions). Where not explicitly available, the \NNNLOgen/\NNLOgen K-factors are parametrised as a superposition of both \NLOgen and \NNLOgen K-factors, allowing the fit to determine the combination of shapes and an overall magnitude. The result is a reduction in $\chi^2$ for the global fit greater than that expected by the extra degrees of freedom.

Recently, there have been new iterations of the three main global PDFs (CT18, MSHT20, NNPDF3.1/4.0). A benchmarking exercise has been conducted for these PDFs, and a combination (PDF4LHC21~\cite{Ball:2022hsh}) has been formed, using Monte Carlo replicas generated from each of the three PDF sets. PDF4LHC21 PDF sets are available either in a 40 member Hessian format, or a 100 member Monte Carlo replica format. The PDF4LHC21 PDFs show a reduction in uncertainty from the combined PDFs determined in 2015, but perhaps not to the extent that may have been expected through the introduction of a variety of new LHC data. This is partially due to the central values of the three input PDFs not coinciding exactly, and partially because the tensions between the data sets that limit the resultant possible uncertainty.%
\footnote{Ref.~\cite{Courtoy:2022ocu} points out one problem that PDF fits may face is the bias that results from improper sampling in very large data spaces. The bias can not only result in an underestimate of the true uncertainty, but also an incorrect central PDF.}
The PDF4LHC21 PDF sets are appropriate for use in general predictions for state-of-the-art calculations, and indeed the prior PDF4LHC15 PDFs have been used in just that way.

\subsection{Development in amplitude and loop integral techniques}

Computing fixed-order amplitudes for scattering processes remains one of the key challenges and obstacles to producing precise predictions for the LHC.
In this section we will give a partial introduction to the procedure and review a selection of some of the most interesting recent advances in this area.
For the purpose of this presentation, we can divide the computation of multi-loop amplitudes into two broad categories: 
\begin{enumerate}
\item Obtaining the amplitudes and simplifying (\textit{reducing}) them,
\item Calculating the integrals which appear in the amplitudes.
\end{enumerate}
They are discussed in Section~\ref{sec:amplitudes} and~\ref{sec:integrals}, respectively.
A thorough review of recent formal developments can be found in Ref.~\cite{Travaglini:2022uwo}. 

Before discussing recent advances in more detail, let us first recall the basic steps.
Firstly, the amplitude is generated either using Feynman diagrams or some other method, for example on-shell techniques.
Next, the tensor structure of the amplitude is separated from the Feynman integrals appearing in the calculation.
The standard approach is known as the \textit{projector method}, the amplitude is expressed as a linear combination of all potential $D$-dimensional tensor structures each multiplied by a scalar coefficient (form factor) containing scalar integrals. 
Projectors are then constructed which, when contracted with the amplitude, pick out the scalar coefficients.
After this, the amplitude may contain a very large number of scalar integrals.
However, not all of these integrals are linearly independent, relations may be found using integration-by-parts identities (IBPs)~\cite{Tkachov:1981wb,Chetyrkin:1981qh} and Lorentz-invariance identities (LIs)~\cite{Gehrmann:1999as}.
An algorithm for eliminating the linearly dependent integrals using these identities was proposed by Laporta~\cite{Laporta:2000dsw} and is now almost ubiquitously used.
At this point, each form factor of an $L$-loop, $n \rightarrow m$-point scattering amplitude can be expressed as
\begin{align}
A^{(L)}_{n\to m} = \sum_j C_j F_j,
\end{align}
where $C_j$ are rational coefficients and $F_j$ are scalar Feynman integrals.
The amplitude has been expressed in terms of fewer (and usually simpler) \textit{master integrals}, the evaluation of which is one of the biggest obstacles in obtaining multi-loop/multi-leg amplitudes.

For many state-of-the-art calculations, every step of the above procedure results in enormously complicated expressions which almost exhaust the available computational resources and take considerable time, see \eg Ref.~\cite{Cordero:2022gsh} for a review.
However, it is often the case that the resulting amplitude is vastly simpler than any intermediate expression encountered during its computation.
This phenomenon is well known in computer science and is called \textit{intermediate expression swell}.
Thus, many recent breakthroughs essentially consist of finding ways to circumvent handling these large intermediate expressions.
For example, the number of Feynman diagrams increases factorially with increasing loop order, they are also a very redundant and gauge dependent representation of the amplitude.
The projector method, especially when applied in $D$ space-time dimensions, can split the amplitude into more form factors than helicity amplitudes present in 4 dimensions.
The set of IBP or LI identities is often over complete and may introduce a large number of integrals into the linear system which do not appear in the amplitude and which are not master integrals.
The goal, then, is primarily finding smart and/or physical ways of avoiding these intermediate complexities and writing down the amplitude in a compact, easy to evaluate form.

\subsubsection{Amplitudes, generalised unitarity and reduction}
\label{sec:amplitudes}

At one-loop, the complexity of high multiplicity processes and the gauge redundancy has greatly been reduced using on-shell and recursive off-shell methods.
These breakthroughs have led to the development of a wide variety of, now commonly, used automated one-loop codes~\cite{Berger:2008sj,Bevilacqua:2011xh,Cascioli:2011va,Buccioni:2019sur,Badger:2012pg,Cullen:2011ac,Cullen:2014yla,Alwall:2014hca,Frederix:2018nkq,Actis:2012qn,Actis:2016mpe,Denner:2017wsf}.

One complexity of the traditional projector method stems from the fact that external particles are handled in $D$ space-time dimensions.
However, for on-shell/physical processes, we can restrict the space-time dimension of the external particles to $D=4$ and directly construct projectors for individual helicity amplitudes. 
In Refs.~\cite{Chen:2019wyb,Peraro:2019cjj,Peraro:2020sfm}, concrete procedures for constructing such helicity projectors were proposed.
The helicity amplitudes obtained using $4$-dimensional projectors are often considerably simpler than the traditional ($D$-dimensional) form factors.
These techniques are now used quite widely in the literature, for example, in the computation of the 2-loop $\gamma \gamma \gamma$~\cite{Chawdhry:2020for} and $\gamma \gamma j$~\cite{Agarwal:2021vdh} amplitudes,
the 3-loop di-photon~\cite{Caola:2020dfu,Bargiela:2021wuy}, gluon-scattering~\cite{Caola:2021izf}, and four-quark~\cite{Caola:2021rqz} amplitudes
as well as several other recent computations, see \eg Refs.~\cite{Bonetti:2020hqh,Chen:2020gae,Badger:2022ncb,Chen:2022rua}.

A useful technique for avoiding intermediate expression swell is the use of \textit{finite fields}, typically integers modulo some prime number $p$, we use $\mathbb{Z}_p$ to denote such a finite field.
Finite fields have long been used for this purpose in computer algebra systems and have thus been implicitly employed in many fixed order computations, for example, via the use of the \textsc{fermat} program~\cite{fermat}.
At the heart of this technique is the concept of black box interpolation.
Consider a univariate function that is known to be a polynomial over the integers $\mathbb{Z}[x]$ (i.e. all coefficients are integer), imagine that we have no knowledge of what this polynomial is but that we have an algorithm to evaluate it on integer input.
By sampling the function on a set of distinct interpolation points $y_0,\ldots,y_d \in \mathbb{Z}$ the coefficients $a_i \in \mathbb{Z}$ of the Newton polynomial
\begin{align*}
f(x) = a_0 + (x-y_0) \left( a_1 + (x-y_1) \left( \ldots + (z - y_{d-1}) a_{d} \right) \right)
\end{align*}
can be recursively reconstructed starting from $f(y_0)$.
The above algorithm can also be recursively applied to reconstruct multivariate polynomials. 
One problem is that intermediate integers (or rational numbers) appearing in the reconstruction can be large and their manipulation (\eg computing greatest common divisors) can be costly.
Instead, by restricting the interpolation points to a finite field $y_0, \ldots, y_d \in \mathbb{Z}_p$ and evaluating $f(x)$ using modular arithmetic, the maximum size of any intermediate expression is limited, largely alleviating the issue of intermediate expression swell.
The reconstruction can be performed using several finite fields $\mathbb{Z}_{p_0},\ldots,\mathbb{Z}_{p_n}$ and the Chinese remainder theorem can be used to deduce $a_i\ (\mathrm{mod}\ p_0 \cdots p_n)$, with enough reconstructions the true value of $a_i \in \mathbb{Z}$ can be determined.
In fact, using the Extended Euclidean Algorithm, it is possible to reconstruct a polynomial over rational numbers $\mathbb{Q}[x]$ from their images modulo integers.
A similar set of considerations apply to rational functions over integers $\mathbb{Z}_p(x)$ and rational functions over rational numbers $\mathbb{Q}_p(x)$, such that they can also be interpolated from samples in finite fields.

An important observation is that finite field methods can be used more broadly in the computation of amplitudes and in their reduction to master integrals.
Some of the earliest direct applications were made in Ref.~\cite{Kant:2013vta}, where they were used to remove redundant equations from the system of IBPs, and in Ref.~\cite{vonManteuffel:2014ixa} where the use of the technique to solve the IBPs was proposed, in Ref.~\cite{Peraro:2016wsq} the finite field techniques were applied to the computation of scattering amplitudes.
Several codes now exist to facilitate the use of finite fields for computing amplitudes, Ref.~\cite{Peraro:2019svx} presents \textsc{FiniteFlow} a \textsc{Mathematica} package that allows every step of the calculation to be encoded in a graph and then performed using finite fields. 
The program {\sc FireFly}~\cite{Klappert:2019emp} provides a \textsc{C++} implementation of many finite field algorithms and is commonly used in conjunction with the \textsc{Kira}~\cite{Maierhofer:2017gsa,Klappert:2020nbg} program for IBP reduction.
In Ref.~\cite{Abreu:2020xvt} the code \textsc{Caravel}, a framework for computing multi-loop scattering amplitudes using numerical unitarity was presented, it supports evaluation using floating point or finite field arithmetic.
Some recent examples of processes computed using finite field methods include the two-loop leading colour helicity amplitudes for: $pp \rightarrow W \gamma + j$~\cite{Badger:2022ncb}, $pp \rightarrow H b\bar{b}$~\cite{Badger:2021ega}, $gg \rightarrow t\bar{t}$~\cite{Badger:2021owl}, $u\bar{d} \rightarrow W b\bar{b}$~\cite{Badger:2021nhg}, as well as the two-loop virtual corrections to $gg \rightarrow \gamma \gamma +j$~\cite{Badger:2021imn} and the two-loop four graviton scattering amplitudes~\cite{Abreu:2020lyk}.
In Ref.~\cite{Laurentis:2019bjh}, a related method for extracting analytic expressions from high-precision floating point evaluations was introduced. 
This technique has been used to compute analytic one-loop amplitudes for $H+4j$~\cite{Budge:2020oyl}, the $pp \rightarrow W(\rightarrow l \nu) + \gamma$~\cite{Campbell:2021mlr} process and $q \bar{q} l \bar{l} l^\prime \bar{l}^\prime g$~\cite{Campbell:2022qpq}.
In Ref.~\cite{DeLaurentis:2022otd}, a related approach to reconstructing analytic expressions from evaluations using $p$-adic numbers was presented.

The use of IBP reduction identities~\cite{Tkachov:1981wb,Chetyrkin:1981qh,Laporta:2000dsw}, LIs~\cite{Gehrmann:1999as}, and dimension shift relations~\cite{Tarasov:1996br,Lee:2009dh} remains a critically important tool in modern loop calculations, but can also present a major bottleneck (see \eg~\cite{Zhang:2016kfo,Grozin:2011mt} for a review).
Several efficient codes exist to facilitate their use, including:
{\sc Air}~\cite{Anastasiou:2004vj}, 
{\sc Fire}~\cite{Smirnov:2008iw,Smirnov:2013dia,Smirnov:2014hma,Smirnov:2019qkx}, 
{\sc LiteRed}~\cite{Lee:2012cn,Lee:2013mka},
{\sc Reduze}~\cite{Studerus:2009ye,vonManteuffel:2012np}
and
{\sc Kira}~\cite{Maierhofer:2017gsa,Maierhofer:2018gpa}.
The use of finite field techniques, as implemented in {\sc FireFly}~\cite{Klappert:2019emp}, {\sc FiniteFlow}~\cite{Peraro:2019svx} and various private codes, has widely been adopted to speed up the reduction to master integrals.

A promising approach to reducing Feynman integrals, without directly solving the system of IBPs, is the use of intersection theory~\cite{Mizera:2017rqa,Mastrolia:2018uzb,Mizera:2019gea,Mizera:2019vvs,Frellesvig:2019uqt,Frellesvig:2019kgj,Abreu:2019wzk,Weinzierl:2020xyy,Frellesvig:2020qot,Caron-Huot:2021xqj,Caron-Huot:2021iev,Chen:2022lzr,Chestnov:2022alh}.
Rather than generating a large set of linear relations between Feynman integrals and solving this system algorithmically, intersection theory instead allows an inner product between pairs of Feynman integrals to be defined.
A basis of preferred Feynman integrals can then be selected (\ie master integrals are chosen) and then each Feynman integral in the problem can be projected onto this basis, with coefficients given by intersection numbers. 
This procedure effectively side-steps the need for IBP reduction.
Unfortunately, as of writing, the computation of (especially multivariate) intersection numbers is itself a computationally expensive process.\footnote{This is an example of the law of conservation of misery.}
For an introduction to the use of intersection theory for Feynman integrals we refer readers to Refs.~\cite{Abreu:2022mfk,Weinzierl:2022eaz}.

Another observation is that often the coefficients of the master integrals can be expressed in a simpler form using multivariate partial fractions, rather than bringing the entire coefficient over a common denominator.
Procedures for performing this partial fraction decomposition, without introducing spurious denominators, were discussed in Refs.~\cite{Bendle:2019csk,Heller:2021qkz}.

Recently, several authors have also examined the use of neural networks to accelerate various aspects of amplitude computation and evaluation.
For example, the use of neural networks to efficiently evaluate high multiplicity and multi-loop amplitudes has been studied in Refs.~\cite{Bishara:2019iwh,Badger:2020uow,Buckley:2020bxg,Aylett-Bullock:2021hmo,Maitre:2021uaa}.
A regularly updated review of the various applications of machine learning in high-energy physics can be found in Ref.~\cite{Feickert:2021ajf}.

\subsubsection{Loop integrals}
\label{sec:integrals}

A modern introduction to various techniques for computing multi-loop Feynman integrals can be found in Ref.~\cite{Weinzierl:2022eaz}, and further details on recent developments can be found in the SAGEX review~\cite{Abreu:2022mfk,Blumlein:2022zkr}.

The use of the differential equations technique~\cite{Kotikov:1990kg,Gehrmann:1999as},
and particularly Henn's canonical form~\cite{Henn:2013pwa} remains as one of the most important methods for computing Feynman Integrals.
New developments concerning the use of differential equations and their application to cutting edge multi-loop integrals can be found in, e.g., Refs.~\cite{Abreu:2020jxa,Frellesvig:2021hkr,Dlapa:2021qsl,Syrrakos:2020kba,Kardos:2022tpo,Henn:2021cyv,Abreu:2021smk}.
In Ref.~\cite{Papadopoulos:2014lla}, a procedure for introducing an auxiliary dimensionless parameter into the kinematics of a process and deriving differential equations with respect to this parameter, known as the simplified differential equations approach, was described.
This procedure has recently been used to compute \eg the 2-loop planar~\cite{Canko:2020ylt} and non-planar~\cite{Kardos:2022tpo} 5-point functions with one massive leg.
For a review of the method of differential equations we refer to Refs.~\cite{Argeri:2007up,Henn:2014qga}.

For a certain class of Feynman integrals, namely finite and \emph{linearly reducible} integrals, the direct integration over the Feynman parameters has proved to be an extremely useful method.
The {\sc HyperInt}~\cite{Panzer:2014caa} package automates this procedure.
This method has been used in several recent calculations, some highlights include the 2-loop mixed QCD-EW corrections to $pp \rightarrow Hg$~\cite{Bonetti:2020hqh,Bonetti:2022lrk}, the quark and gluon form factors at four-loops in QCD~\cite{Lee:2021uqq,Lee:2022nhh} and the $Hb\bar{b}$ vertex at four-loops~\cite{Chakraborty:2022yan}.

Many scattering amplitudes computed in the last few decades can be expressed in terms of multiple polylogarithms (MPLs). 
The mathematical properties of these functions are well understood and public tools exist for their numerical evaluation.
However, it has long been established that not all Feynman integrals can be expressed in terms of MPLs.
For example, such integrals appear in $\gamma \gamma$, $t \bar{t}(+j)$, $t\bar{t}H$ and $H+j$ production as well as Higgs decays and many other processes of interest at the LHC.
One obstruction is the appearance of elliptic or hyperelliptic integrals.
An enormous effort is ongoing to analytically tackle such integrals and generalise the concepts and tools previously developed for MPLs.
For a recent review of the various developments, we refer the reader to Ref.~\cite{Bourjaily:2022bwx}.

For some processes, it is not straightforward to obtain a compact and easy to evaluate analytic expression for the Feynman integrals.
In these cases, it is often convenient to use a numerical or semi-numerical method for their evaluation.
One method, that has recently been systematised and used widely for the evaluation of multi-scale Feynman integrals, is the use of generalised power series.
The key insight is that, given the differential equations for a set of master integrals, $\vec{F}$, with respect to some variable $t$ (which parametrises a contour $\gamma(t)$)
\begin{align}
\frac{\partial}{\partial t} \vec{F}(t,\epsilon) = \mathbf{A}_t(t,\epsilon) \vec{F}(t,\epsilon),
\end{align}
and the value of the integrals at some point $t_0$ on the path, series solutions can be obtained both in the vicinity of (regular) singular points (using the Frobenius method) and in the vicinity of regular points (using Taylor expansion).
The contour $\gamma(t)$ can be chosen such that solutions of the master integrals at one phase-space point (or for particular values of the masses) can be transported along the contour to another phase-space point by matching series expansions around regular and/or singular points along the path.
This method was recently described in detail and popularised in Refs.~\cite{Moriello:2019yhu,Bonciani:2019jyb,Frellesvig:2019byn} and has been implemented in the code \textsc{DiffExp}~\cite{Hidding:2020ytt}.
An obstruction to the use of this method is the requirement to know a priori the value of the integrals at the point $t_0$, \ie the boundary conditions.
One option is to obtain these boundary conditions numerically in the Euclidean region, for example by using sector decomposition, this approach was applied to 2- and 3-loop integrals in Ref.~\cite{Dubovyk:2022frj}.
In Refs.~\cite{Liu:2017jxz,Liu:2020kpc,Liu:2021wks}, an innovative solution to this problem, now known as the auxiliary mass flow (AMF) method, was proposed.
The Feynman propagators appearing in loop integrals can be generically be written as
\begin{align}
\frac{1}{k^2-m^2+i \eta},
\end{align}
where $k$ is the momentum of the propagating particle, $m$ is the mass and $\eta$ is an infinitesimal parameter used to define the causal integration contour, ultimately we set $\eta \rightarrow 0_+$.
It was noticed that if one instead chooses $\eta = \infty$, this corresponds to a large mass and it is straightforward to obtain boundary conditions here via a large mass expansion, the contour $\gamma(t)$ can then be chosen along the $\eta$ direction connecting the point at $\eta = \infty$ to a physical phase-space point at $\eta = 0_+$.
Tools for using the AMF method have recently been implemented in the package \textsc{AMFlow}~\cite{Liu:2022chg}
This technique has been used to calculate amplitudes for $gg \rightarrow WW$~\cite{Bronnum-Hansen:2020mzk}, $gg \rightarrow ZZ$ and ~\cite{Bronnum-Hansen:2021olh}, t-channel single top production~\cite{Bronnum-Hansen:2021pqc,Bronnum-Hansen:2022tmr} and integrals relevant for a variety of processes at 2- and 3-loops~\cite{Liu:2021wks}.

Very recently, a method that iteratively combines propagators using ``Feynman's trick'' and then solves the resulting Feynman parameter integrals in terms of generalised series expansions, using an associated system of simplified differential equations, was described in Ref.~\cite{Hidding:2022ycg}.
This method was used to numerically obtain results for a two-loop non-planar double pentagon family with 40 digits of precision.

Direct numerical evaluation of Feynman integrals can also be a useful technique.
It is an especially promising strategy for tackling multi-loop integrals with many internal masses and where analytic results can not straightforwardly be obtained.
The sector decomposition algorithm~\cite{Binoth:2000ps} has seen a number of optimisations, implemented into the publicly available
updates of the codes {\sc (py)SecDec}~\cite{Borowka:2015mxa,Borowka:2017idc,Borowka:2018goh,Heinrich:2021dbf} and {\sc Fiesta}~\cite{Smirnov:2015mct,Smirnov:2021rhf}.
By numerically integrating in Feynman parameter space, the virtual amplitude for $gg \to ZZ$~\cite{Agarwal:2020dye} and $pp\to ZH$~\cite{Chen:2020gae,Chen:2022rua} at NLO including the full top quark mass dependence have been completed.
Recently, in Ref.~\cite{Winterhalder:2021ngy}, a method for finding an optimal integration contour for numerical integration using neural networks was explored.
In Ref.~\cite{Borinsky:2020rqs} a method for numerically evaluating Feynman integrals using tropical geometry was shown to be very efficient for a special class of (Euclidean) integrals.

Most of the approaches above rely on the explicit separation of real and virtual corrections. 
Instead, in the Loop--Tree Duality framework, these contributions are treated together, which can help to avoid having to separately treat the IR divergences arising in and then cancelling between the amplitudes.
Progress continues to be made in this direction with several recent works continuing to develop and automate the technique as well as demonstrating its use for various higher-order calculations~\cite{Driencourt-Mangin:2019yhu,Capatti:2020xjc,Prisco:2020kyb,Sborlini:2021owe,TorresBobadilla:2021ivx,Bobadilla:2021pvr,Kermanschah:2021wbk,Capatti:2022tit}.
In Refs.~\cite{RamirezUribe:2021nlj, Ramirez-Uribe:2021ubp}, the utility of Grover's algorithm (a quantum computing algorithm) for Loop--Tree Duality was highlighted.
We also point the reader to a recent review~\cite{deJesusAguilera-Verdugo:2021mvg}.

\subsection{Infrared subtraction methods for differential cross sections}

Differential higher-order calculations must retain the full information on the final-state kinematics, which also includes the regions of real-emission phase space that are associated with soft and/or collinear configurations.
While the associated singularities must cancel with the explicit poles in the virtual amplitudes for any infrared (IR) safe observable, this entails some form of integration of the unresolved emission to expose the singularity. 
IR subtraction methods facilitate the explicit cancellation of singularities to obtain finite cross sections,
\begin{equation}
  d\sigma_{2\to n} \text{\NLO{k}} = {\rm IR}_k(A^{k}_{2\to n}, A^{k-1}_{2\to n+1},\cdots, A^{0}_{2\to n+k})\,,
\end{equation}
where the function ${\rm IR}_k$ represents an infrared subtraction technique that leaves the kinematic information for each particle multiplicity intact.

While full automation of \NLOgen subtractions has been achieved, this is not yet the case at \NNLOgen.
Nonetheless, tremendous progress has been made in differential \NNLOgen calculations, essentially completing all relevant $2 \to 1$ and $2 \to 2$ processes. 
This puts the next frontier in \NNLOgen calculations to $2 \to 3$ processes, as well as revisiting prior approximations that could potentially limit the interpretation of theory--data comparisons (\eg combination of production and decay subprocesses, flavoured jet definition, photon-jet separation and hadron fragmentation, on-shell vs.\ off-shell, etc.).
Lastly, we have observed remarkable progress in the area of differential \NNNLOgen calculations with first results obtained for $2 \to 1$ benchmark processes.

\begin{itemize}
\item Antenna subtraction~\cite{Gehrmann-DeRidder:2005btv,Currie:2013vh}:\\
  Applicable to processes with hadronic initial and final states with analytically integrated counterterms. 
  An almost completely local subtraction up to angular correlations that are removed through the averaging over azimuthal angles. 
  Applied to processes in \(\mathrm{e}^+\mathrm{e}^-\), deep-inelastic scattering (DIS), and hadron–hadron collisions:
  \(\mathrm{e}^+\mathrm{e}^-\to 3j\)~\cite{Gehrmann-DeRidder:2014hxk,Gehrmann:2017xfb},
  (di-)jets in DIS~\cite{Currie:2017tpe,Niehues:2018was},
  \(pp\to \text{(di)-jets}\)~\cite{Currie:2016bfm,Currie:2017eqf},
  \(pp\to \gamma\gamma\)~\cite{Gehrmann:2020oec},
  \(pp\to \gamma+j/X\)~\cite{Chen:2019zmr},
  \(pp\to V+j\)~\cite{Gehrmann-DeRidder:2015wbt,Gehrmann-DeRidder:2016cdi,Gehrmann-DeRidder:2017mvr},
  \(pp\to H+j\)~\cite{Chen:2016zka},
  \(pp\to VH(+\mathrm{jet})\)~\cite{Gauld:2019yng,Gauld:2020ced,Gauld:2021ule},
  and Higgs production in VBF~\cite{Cruz-Martinez:2018rod}.
  Extensions to cope with identified jet flavours~\cite{Gauld:2019yng,Gauld:2020deh} and the photon fragmentation function~\cite{Gehrmann:2022cih,Chen:2022gpk}.

\item Sector-improved residue subtraction~\cite{Czakon:2010td,Czakon:2011ve,Boughezal:2011jf}:\\
  Capable of treating hadronic initial and final states through a fully local subtraction that incorporates ideas of the FKS approach at NLO~\cite{Frixione:1995ms,Frederix:2009yq} and a sector decomposition~\cite{Binoth:2000ps} approach
  for real radiation singularities~\cite{Heinrich:2002rc,Anastasiou:2003gr,Binoth:2004jv}.
  Counterterms obtained numerically with improvements using a four-dimensional formulation~\cite{Czakon:2014oma}.
  Applied to
  top-quark processes~\cite{Czakon:2013goa,Czakon:2014xsa,Czakon:2015owf,Czakon:2016ckf,Brucherseifer:2013iv,Brucherseifer:2014ama}, to $pp\to H+j$~\cite{Boughezal:2015dra,Caola:2015wna}, inclusive jet production~\cite{Czakon:2019tmo}, $pp\to3\gamma$~\cite{Chawdhry:2019bji},
  $pp\to2\gamma+j$~\cite{Chawdhry:2021hkp},
  $pp\to3j$~\cite{Czakon:2021mjy},
  $pp\to W+j$~\cite{Pellen:2022fom}.
  Extensions to deal with flavoured jets~\cite{Czakon:2020coa} and $B$-hadron production~\cite{Czakon:2021ohs}.
  
\item $q_T$-subtraction~\cite{Catani:2007vq}:\\
  A slicing approach for processes with a colourless final state and/or a pair of massive coloured particles.
  Applied to
  $H$~\cite{Catani:2007vq,Grazzini:2008tf},
  $V$~\cite{Catani:2009sm,Catani:2010en}
  and $VV'$ production processes~\cite{Catani:2011qz,Grazzini:2013bna,Gehrmann:2014fva,Cascioli:2014yka,Grazzini:2015nwa,Grazzini:2015hta,Grazzini:2016swo,Grazzini:2016ctr,Grazzini:2017ckn,Catani:2018krb,Kallweit:2018nyv}, which are available in the \Matrix program~\cite{Grazzini:2017mhc}.
  Predictions at \NNLOQCD for $H$, $V$, $VH$, $V\gamma$, $\gamma\gamma$, and $VV'$ available in the MCFM program~\cite{Campbell:2022gdq}.
  Further applications at \NNLOQCD include $VH$~\cite{Ferrera:2011bk,Ferrera:2013yga,Ferrera:2014lca}, $HH$~\cite{deFlorian:2016uhr,Grazzini:2018bsd}, $VHH$~\cite{Li:2016nrr,Li:2017lbf}.
  Extended to cope with a pair of massive coloured particles~\cite{Bonciani:2015sha,Angeles-Martinez:2018mqh} and applied to top-pair production~\cite{Catani:2019iny,Catani:2019hip} and $b\bar{b}$ production~\cite{Catani:2020kkl}.
  The same developments allowed the mixed QCD-EW corrections to Drell-Yan with massive leptons to be tackled~\cite{Buonocore:2021rxx,Bonciani:2021zzf}.
  Method extended to \NNNLOQCD with applications to Higgs production~\cite{Cieri:2018oms,Billis:2021ecs} and Drell-Yan production~\cite{Chen:2021vtu,Camarda:2021ict,Camarda:2021jsw,Chen:2022cgv,Chen:2022lwc}.
  
\item $N$-jettiness~\cite{Boughezal:2015eha,Boughezal:2015dva,Gaunt:2015pea}:\\
  A slicing approach based on the resolution variable $\tau_N$ ($N$-jettiness) that is suited for processes beyond the scope of the $q_T$ method, i.e.\ involving final-state jets. 
  Explicitly worked out at \NNLOQCD for hadron-collider processes with up to one jet.
  Applied to $V(+j)$~\cite{Boughezal:2015dva,Boughezal:2015aha,Boughezal:2015ded,Boughezal:2016isb,Boughezal:2016yfp,Boughezal:2016dtm,Campbell:2016jau,Campbell:2016lzl,Campbell:2017aul} and
  $H+j$~\cite{Campbell:2019gmd}.
  Colourless final state production available in the MCFM program~\cite{Boughezal:2016wmq,Campbell:2019dru}.
  Same technique also used in the calculation of top decay~\cite{Gao:2012ja} and $t$-channel single top production~\cite{Berger:2016oht}.
  Important steps towards the extension for \NNNLOQCD calculations have been made in Refs.~\cite{Melnikov:2018jxb,Melnikov:2019pdm,Behring:2019quf,Billis:2019vxg,Baranowski:2020xlp,Ebert:2020unb,Baranowski:2022khd}.

\item ColorFul subtraction~\cite{DelDuca:2015zqa}:\\
  Fully local subtraction extending the ideas of the Catani--Seymour dipole method at NLO~\cite{Catani:1996vz}.
  Analytically integrated counter-terms for the infrared poles, numerical integration for finite parts.
  Fully worked out for processes with hadronic final states and applied to $H\to b\bb$~\cite{DelDuca:2015zqa} and $e^+e^-\to$ 3 jets~\cite{DelDuca:2016csb,DelDuca:2016ily,Tulipant:2017ybb}.

\item Nested soft--collinear subtraction~\cite{Caola:2017dug,Caola:2018pxp,Delto:2019asp}:\\
  Fully local subtraction with analytic results for integrated subtraction counterterms. 
  Worked out for processes with hadronic initial and final states~\cite{Caola:2019nzf, Caola:2019pfz,Asteriadis:2019dte}.
  Applied to compute \NNLOQCD corrections to VH~\cite{Caola:2017xuq} and VBF~\cite{Asteriadis:2021gpd}, as well as mixed QCD-EW corrections to the Drell-Yan process~\cite{Delto:2019ewv,Buccioni:2020cfi,Behring:2020cqi}.

\item Local analytic sector subtraction~\cite{Magnea:2018hab, Magnea:2018ebr,Magnea:2020trj}:\\
  Local subtraction with analytic integration of the counterterms aiming to combine the respective advantages from two NLO approaches of FKS subtraction~\cite{Frixione:1995ms,Frederix:2009yq} and dipole subtraction~\cite{Catani:1996vz}.
  Fist proof-of-principle results for $e^+e^-\to 2$\,jets~\cite{Magnea:2018hab}.

\item Projection to Born~\cite{Cacciari:2015jma}:\\
  Requires the knowledge of inclusive calculations that retain the full differential information with respect to Born kinematics. 
  With the necessary ingredients in place, generalisable to any  order.
  Applied at \NNLOQCD to VBF~\cite{Cacciari:2015jma}, Higgs-pair production~\cite{Dreyer:2018rfu}, and $t$-channel single top production~\cite{Berger:2016oht,Campbell:2020fhf}.
  Fully differential \NNNLOQCD predictions obtained for jet production in DIS~\cite{Currie:2018fgr,Gehrmann:2018odt}, $H\to b\bar{b}$~\cite{Mondini:2019gid}, and Higgs production in gluon fusion~\cite{Chen:2021isd}.
\end{itemize}


\section{Update on the precision Standard Model wish list}
\label{sec:SM_wishlist:precision_wish_list}

The summary is broken up into four different parts which comprise: Higgs processes, jet production, associated vector-boson production, and top-quark processes.

The perturbative corrections are defined with respect to the leading order prediction in QCD and the expansion with respect to the strong and electroweak couplings read as:
\begin{equation}
  d\sigma_X = d\sigma_X^{\rm LO} \left(1 +
      \sum_{k=1} \alpha_s^k d\sigma_X^{\delta \text{\NLOQ{k}}}
    + \sum_{k=1} \alpha^k d\sigma_X^{\delta \text{\NLOE{k}}}
    + \sum_{k,l=1} \alpha_s^k \alpha^l d\sigma_X^{\delta \text{\NLOQE{k}{l}}}
    \right).
  \label{eq:SM_wishlist:dsigmapertexp}
\end{equation}
The mixed QCD--EW corrections are singled out to distinguish between additive predictions QCD+EW and mixed predictions QCD$\otimes$EW. 
Equation~\eqref{eq:SM_wishlist:dsigmapertexp} only applies to cases where the leading-order process is uniquely defined.
For cases with multiple types of tree level amplitudes (requiring at least two jets at hadron colliders), it is customary to classify the Born process as the one with the highest power in $\alpha_s$ that is typically the dominant contribution.
In the following, the notation \NLOSM is used to denote NLO calculations that include the complete Standard Model corrections, \ie all QCD and EW corrections to all leading-order contributions.

Tremendous progress has been made in the field of resummation and parton showers, however, we refrain from reviewing it here and instead point the interested reader to Ref.~\cite{Campbell:2022qmc}.
The present authors feel that this rich area of research certainly deserves a dedicated summary and wishlist of its own.%
\footnote{The Les Houches-Wishlist wishlist.}

Below, a summary of the current status of higher-order computations within the Standard Model is provided.
The references retained are for the most state-of-the-art calculations available at the time of submission.
In that regards, superseded computations are not listed here.
Specifically, we provide a brief summary of the status of theory predictions as documented in the previous wishlist (LH19), followed by a description of the progress since then.
A short experimental motivation for the entries in the wishlist is given with the anticipated precision target that should be achieved.
Before turning to the actual wishlist, we discuss some general aspects pertaining to higher-order calculations at the LHC.

\paragraph*{Electroweak corrections}

A naive estimate based solely on the size of the respective coupling constants, $\alpha \sim \alpha_s^2$, already highlights that \NLOEW corrections should be considered together with \NNLOQCD for any applications aiming at the per-cent level precision. 
Moreover, EW corrections can be enhanced in specific kinematic regimes such as high-energy tails of distributions that are prone to large logarithmic corrections, known as EW Sudakov logarithms, or observables that are sensitive to final-state radiation effects. 
In such scenarios, \NLOEW corrections can compete with or even surpass the size of \NLOQCD corrections and are essential to be included in the theory predictions.
For benchmark processes such as the Drell--Yan process, also \NNLOgen mixed QCD--EW corrections must be considered to match the precision targets of the experimental measurements.

Recent years have seen immense progress in the automation of \NLOEW corrections, with dedicated one-loop Matrix Element providers such as
\OpenLoops~\cite{Cascioli:2011va,Buccioni:2019sur}, \GoSam~\cite{Cullen:2011ac,Cullen:2014yla,Chiesa:2017gqx}, \Recola~\cite{Actis:2012qn,Actis:2016mpe,Denner:2017wsf}, \MadLoop~\cite{Alwall:2014hca,Frederix:2018nkq}, and \NLOX~\cite{Honeywell:2018fcl},
enabling the calculation of \NLOEW corrections to very complex off-shell processes with multiplicities up to $2\to6$~\cite{Denner:2016jyo,Biedermann:2017bss,Denner:2019tmn,Denner:2020zit,Denner:2021hsa,Denner:2022pwc} and even $2\to7$~\cite{Denner:2016wet}.
The universal EW Sudakov logarithms have further been incorporated into automated frameworks~\cite{Bothmann:2020sxm,Pagani:2021vyk} based on Ref.~\cite{Denner:2000jv}.

These advances in fixed-order EW computations are therefore naturally accompanied by the need of including them in global PDF fits.
The LUXqed~\cite{Manohar:2016nzj,Manohar:2017eqh} approach enabled the precise determination of the photon PDF and the extension of interpolation grid technologies to cope with arbitrary perturbative corrections (QCD, EW, or mixed)~\cite{Carrazza:2020gss} paves the way for a consistent inclusion of EW effects into global PDF fits. 
Beyond fixed-order corrections, several attempts have been made to combine these with QCD and/or EW parton-showers.
For the latter case, there has been dedicated progress in incorporating QED/EW effects in parton showers~\cite{Skands:2020lkd,Kleiss:2020rcg,Gutschow:2020cug,Masouminia:2021kne}.
In the future, one can hope to have full EW corrections matched to parton/EW showers for arbitrary processes.
For several exemplary processes~\cite{Barze:2013fru,Granata:2017iod,Chiesa:2020ttl}, an exact QED matching along with QCD corrections have been already obtained at full NLO accuracy (EW + QCD).

Another aspect of EW corrections that has seen recent progress concerns the definition of PDF and fragmentation functions.
For example, there has been substantial work in determining the photon~\cite{Manohar:2016nzj,Manohar:2017eqh} and lepton~\cite{Buonocore:2020nai,Buonocore:2020erb,Buonocore:2021bsf} content of the proton.
Similarly, the quark and gluon content of leptons has been investigated using various approaches~\cite{Han:2021kes,Frixione:2012wtz,Frixione:2019lga}.
The photon-to-jet conversion function has also been derived in Ref.~\cite{Denner:2019zfp} in order to deal with EW corrections to processes with jets in the final state.
Related work on the treatment of isolated photons within higher-order EW corrections has been done in Ref.~\cite{Pagani:2021iwa}.

For further details on EW radiative corrections, we refer to the exhaustive review provided in Ref.~\cite{Denner:2019vbn}.

\paragraph*{On-shell and off-shell descriptions}

The treatment of unstable particles constitutes a subtle issue and can be approached at different levels of sophistication.
In the crudest approximation, the massive states, \eg W and Z bosons and the top quark, are considered stable and the production of on-shell states is computed. 
In case of a narrow resonance, the decay can be included in a factorised manner through the so-called narrow-width approximation (NWA), where the intermediate propagator is approximated by an on-shell delta distribution.
Several variants of the NWA exist that further incorporate spin correlations between the production and decay sub-processes or include some finite-width effects through the resonance shape. 
The intrinsic error of this approximation is of the order $\mathcal{O}\left(\Gamma_i/M_i\right)$, with $\Gamma_i$ and $M_i$ the width and the mass of the resonant particle, respectively.
This estimate holds provided the decay products are treated inclusively and the resonant contributions dominate~\cite{Fadin:1993kt,Fadin:1993dz,Melnikov:1993np,Uhlemann:2008pm}.

An improved treatment can be obtained using the pole approximation (PA)~\cite{Bardin:1988xt,Stuart:1991xk,Aeppli:1993rs,Beenakker:1998gr,Denner:2000bj}, which performs a systematic expansion about the resonance pole.
In this case, the resonant propagator(s) are kept intact, while the residue is evaluated at the on-shell point.
The full phase-space kinematics can be accommodated together with an on-shell projection and spin correlations between production and decay can be implemented.
This approximation has the advantage to describe the full resonant shape without reverting to an expensive off-shell computation.
It can also be used to infer the size of non-resonant contributions by comparing it to a full off-shell computation.
Finally, the PA is particularly suited for polarised predictions given that these require intermediate on-shell states to define the corresponding polarisation.

Lastly, off-shell calculations refer to predictions that include all resonant and non-resonant diagrams that contribute to a given final state as defined by the decay products.
The drawback of such predictions is that they are significantly more complicated than the previous approximations and therefore more CPU intensive.
While multiplicities up to $2\to6/7/8$ have been achieved at \NLOgen accuracy, \NNLOQCD calculations are only starting to break into the $2\to3$ barrier.

In the following, off-shell effects are assumed to be taken into account when the status of the wishlist is discussed.
In the case of QCD corrections and purely EW decays, the different treatment of the unstable particles in general does not significantly complicate the calculation.
For calculations where this is not the case, \eg EW corrections and top-quark processes, the off-shell effects are mentioned explicitly.

\paragraph*{Fiducial cuts}

The interplay between fiducial selection cuts and higher-order radiative corrections has received increased attention in the recent years, in particular, in regards to the associated linear fiducial power corrections~\cite{Frixione:1997ks,Ebert:2019zkb,Alekhin:2021xcu}.
These can be fully accommodated into resummed predictions through a simple recoil prescription~\cite{Catani:2015vma,Ebert:2020dfc}. 
In the context of fixed-order calculations based on slicing methods, this further allows to significantly stabilize the variation with the slicing parameter. 
The impact of fiducial power corrections on cross sections can become sizeable as was observed in the gluon-fusion Higgs production process with ATLAS cuts~\cite{Billis:2021ecs}, while in the case of the Drell--Yan process their impact was found to be only of moderate size~\cite{Chen:2022cgv}.
It was shown in Ref.~\cite{Salam:2021tbm} that linear fiducial power corrections can be almost entirely avoided in two-body decay processes through a suitable adjustment of the selection cuts. 
In its simplest incarnation, this can be accomplished by applying cuts on ``self-balancing variables'' such as the arithmetic $\tfrac{1}{2}(p_{T,1}+p_{T,2})$ or geometric $(p_{T,1}\cdot p_{T,2})^{1/2}$ mean of the transverse momenta. Such cuts should not be difficult to implement in most experimental analyses, and will not result in a significant loss in acceptance. 
In the case of the Drell--Yan process, it was seen that this approach indeed largely eliminates the impact from fiducial power corrections and in the same way reduces the difference between resummed and fixed-order predictions on inclusive quantities such as the fiducial cross section.

\paragraph*{Jet algorithms, identified final states, and fragmentation}

\NNLOgen predictions are necessary to achieve the highest precision for $2\to 2$ (and $2\to 3$) processes at the LHC. 
The presence of one or more jets in the final state requires the application of a jet algorithm, almost universally the anti-$k_t$ algorithm as they give rise to geometrically regular jets. 
However, there can be accidental cancellations between the scale-dependent terms in the \NNLOgen calculation that can result in artificially small scale uncertainties, especially close to jet radii of $R=0.4$. 
A more realistic estimate of the uncertainty can be obtained by the use of a larger radius jet ($R=0.6$--$0.7$), or by alternate estimates for uncertainties from missing higher orders~\cite{Bellm:2019yyh,Rauch:2017cfu,Buckley:2021gfw}. 

Increasingly, many of the precision LHC measurements involve the presence of heavy quarks in the final state, e.g.\ V+c/b. 
The heavy flavour quark is reconstructed as a jet with a heavy flavour tag, imposing a transverse momentum threshold that is typically much smaller than the transverse momentum of the jet itself.
Calculations at \NNLOgen require the application of an IR-collinear safe jet algorithm such as the flavour-$k_t$ algorithm~\cite{Banfi:2006hf}. 
The experimental approach, however, is to first reconstruct the jet using the anti-$k_t$ jet algorithm, and then to look for the presence of heavy flavour within that jet.
The resulting mis-match in algorithms can result in an error of the order of 10\%, potentially larger than the other sources of uncertainty in the measurement/prediction.
Such a mismatch can be avoided through a computation based on massive heavy quarks (see e.g.\ Ref.~\cite{Behring:2020uzq} for a comparison against flavour-$k_t$ in $WH$ production) or by the inclusion of the fragmentation contribution at \NNLOgen (see e.g.\ Ref.~\cite{Czakon:2021ohs} for \NNLOQCD predictions for $B$-hadron production in $t\bar{t}$). 
Alternatively, the mismatch can be mitigated through new jet-tagging procedures~\cite{Caletti:2022hnc,Caletti:2022glq,Czakon:2022wam}.

A similar issue with a mismatch between experiment and theory arises in the case of identified photons that require an isolation procedure to distinguish the prompt production from the overwhelming background. 
Differences in a fixed-cone isolation versus a smooth-cone isolation~\cite{Frixione:1998jh,Siegert:2016bre} have been the subject of many studies which assessed the impact to be at the few-percent level~\cite{Andersen:2014efa,Andersen:2016qtm,Catani:2018krb,Catani:2013oma,Amoroso:2020lgh}.
Precision phenomenology based on processes with external photons thus demands for an extension of the fragmentation contribution to \NNLOgen that has been achieved recently~\cite{Gehrmann:2022cih,Chen:2022gpk}.

\subsection{Higgs boson associated processes}

An overview of the status of Higgs boson associated processes is given in Table~\ref{tab:SM_wishlist:wlH}.
In the following, the acronym \emph{Heavy Top limit} (HTL) is used to denote the effective field theory in the $m_t\to\infty$ limit.
In this limit, the Higgs bosons couple directly to gluons via the following effective Lagrangian
\begin{equation}
\mathcal{L}_{\rm eff} = - \frac{1}{4} G^a_{\mu \nu} G_a^{\mu \nu} \left(C_H \frac{H}{v} - C_{HH} \frac{H^2}{2 v^2} + C_{HHH} \frac{H^3}{3 v^3} + \ldots\right)\,,
\end{equation}
whose matching coefficients known up to fourth order in $\alpha_S$~\cite{Chetyrkin:1997iv,Chetyrkin:2005ia,Kramer:1996iq,Schroder:2005hy,Djouadi:1991tka,Grigo:2014jma,Spira:2016zna,Gerlach:2018hen}.

\begin{table}
  \renewcommand{\arraystretch}{1.5}
\setlength{\tabcolsep}{5pt}
  \begin{center}
  \begin{tabular}{lll}
    \hline
    \multicolumn{1}{c}{process} & \multicolumn{1}{c}{known} &
    \multicolumn{1}{c}{desired} \\
    \hline
    $pp\to H$ &
    \begin{tabular}{l}
      \NNNLOHTL \ \\
      \NNLOQCDT \ \\
      \NLOHE11 \\ 
      \NLOQCD
    \end{tabular} &
    \begin{tabular}{l}
      \NLOH4 (incl.) \\
      \NNLOQCDBC 
    \end{tabular} \\
    \hline
    $pp\to H+j$ &
    \begin{tabular}{l}
      \NNLOHTL \\
      \NLOQCD \\
      \NLOQE11 \\ 
    \end{tabular} &
    \begin{tabular}{l}
      \NNLOHTL$\!\otimes\,$\NLOQCD\!+\,\NLOEW
    \end{tabular} \\
    \hline
    $pp\to H+2j$ &
    \begin{tabular}{l}
      \NLOHone$\!\otimes\,$\LOQCD \\
      \NNNLOQCDVBFstar (incl.) \\
      \NNLOQCDVBFstar  \\ 
      \NLOEWVBF 
    \end{tabular} &
    \begin{tabular}{l}
      \NNLOHTL$\!\otimes\,$\NLOQCD\!+\,\NLOEW\\
      \NNNLOQCDVBFstar \\
      \NNLOQCDVBF
    \end{tabular} \\
    \hline
    $pp\to H+3j$ &
    \begin{tabular}{l}
      \NLOHone \\
      \NLOQCDVBF
    \end{tabular} &
    \begin{tabular}{l}
      \NLOQCD\!+\,\NLOEW \\
    \end{tabular} \\
    \hline
    $pp\to VH$ &
    \begin{tabular}{l}
      \NNLOQCD\!+\,\NLOEW \\
      \NLOggHVtb{} \\
    \end{tabular} &
    \begin{tabular}{cl}
    \end{tabular} \\
    \hline
    $pp\to VH + j$ &
    \begin{tabular}{l}
      \NNLOQCD \ \\
      \NLOQCD\!+\,\NLOEW \\
    \end{tabular} &
    \begin{tabular}{cl}
      \NNLOQCD + \NLOEW 
    \end{tabular} \\
    \hline
    $pp\to HH$ &
    \begin{tabular}{l}
      \NNNLOHTL$\!\otimes\,$\NLOQCD \\
    \end{tabular} &
    \begin{tabular}{cl}
      \NLOEW \\
    \end{tabular} \\
    \hline   
    $pp\to HH + 2j$ &
    \begin{tabular}{l}
      \NNNLOQCDVBFstar (incl.) \\
      \NNLOQCDVBFstar  \\ 
      \NLOEWVBF 
    \end{tabular} &
    \begin{tabular}{cl}
      \\
    \end{tabular} \\
    \hline
    $pp\to HHH$ &
    \begin{tabular}{l}
      \NNLOHTL \\
    \end{tabular} &
    \begin{tabular}{cl}
      \\
    \end{tabular} \\
    \hline
    $pp\to H+t\tb$ &
    \begin{tabular}{l}
      \NLOQCD\!+\,\NLOEW\\
      \NNLOQCD (off-diag.) 
    \end{tabular} &
    \begin{tabular}{l}
     \NNLOQCD
    \end{tabular}  \\
    \hline
    $pp\to H+t/\tb$ &
    \begin{tabular}{l}
      \NLOQCD\!+\,\NLOEW\\
    \end{tabular} &
    \begin{tabular}{l}
      \NNLOQCD 
    \end{tabular} \\
    \hline
  \end{tabular}
  \caption{Precision wish list: Higgs boson final states. \NLOQVBFstar{x} means a
   calculation using the structure function approximation. $V=W,Z$.}
  \label{tab:SM_wishlist:wlH}
  \end{center}
\renewcommand{\arraystretch}{1.0}
\end{table}

\begin{itemize}[leftmargin=2cm]

\item[$H$:]

\textit{LH19 status}:
Results at \NNLOHTL known for two decades~\cite{Harlander:2002wh,Anastasiou:2002yz,Ravindran:2003um,Catani:2007vq,Grazzini:2008tf}.
Inclusive \NNNLOHTL results computed in \cite{Anastasiou:2015vya,Anastasiou:2016cez,Mistlberger:2018etf} and available exactly in the program {\sc iHixs 2}~\cite{Dulat:2018rbf} and in an expansion around the Higgs production threshold in {\sc SusHi}~\cite{Harlander:2016hcx}. 
The first differential results at \NNNLOHTL were presented in Ref.~\cite{Dulat:2017brz,Dulat:2017prg,Dulat:2018bfe,Cieri:2018oms} and the transverse momentum spectrum of the Higgs boson has been studied at \NNLOgen\!+\,\NNNLL~\cite{Chen:2018pzu,Bizon:2018foh}.
The $m_t$-dependence is known at 3-loops for the virtual piece~\cite{Davies:2019nhm,Czakon:2020vql,Harlander:2019ioe} and at 4-loops in a large-$m_t$ expansion~\cite{Davies:2019wmk}.
Complete \NLOQCD corrections are known for arbitrary quark masses~\cite{Dawson:1990zj,Djouadi:1991tka,Graudenz:1992pv,Spira:1995rr,Harlander:2005rq,Anastasiou:2006hc,Aglietti:2006tp,Anastasiou:2009kn}.
Bottom quark effects have been studied for intermediate Higgs transverse momentum $m_b \lesssim p_T \lesssim m_t$ at \NLOgen\!+\,\NNLL~\cite{Caola:2018zye}.
Mixed QCD--EW corrections, \NLOHE11, were known in the limit of small electroweak gauge boson mass~\cite{Bonetti:2018ukf,Anastasiou:2018adr}.
\medskip 

The complete \NNNLOHTL corrections  were computed fully differentially, at fixed order~\cite{Chen:2021isd} and with \NNNLLp resummation for the Higgs $p_T$ spectrum~\cite{Billis:2021ecs}.
After applying fiducial cuts, it was observed that the fixed order corrections exhibit some instabilities stemming from linear power corrections $\sim p_{T,H}/m_H$ which are cured by the \NNNLLp resummation~\cite{Billis:2021ecs}.
These instabilities are substantially removed using the cuts suggested in Ref.~\cite{Salam:2021tbm}, as described above.
Ultimately, the resulting \NNNLOHTL and \NNNLLp+\NNNLOHTL corrections are observed to give an enhancement similar to the inclusive case. 
In Ref.~\cite{Re:2021con} fiducial results for Higgs Boson production were produced at \NNNLLp + \NNLOHTL using the RadISH formalism.
A new program, \textsc{Hturbo}, for producing fast \NNLOgen+\NNLL predictions for $gg \rightarrow H(\rightarrow \gamma \gamma)$,  was presented in Ref.~\cite{Camarda:2022wti}, it represents an independent reimplementation of the \textsc{HqT}~\cite{Bozzi:2005wk,deFlorian:2011xf}, \textsc{HNNLO}~\cite{Catani:2007vq} and \textsc{HRes}~\cite{deFlorian:2012mx} programs.

The dominant light-quark contribution to the \NLOgen mixed QCD-EW corrections have now been computed including the exact EW-boson mass dependence~\cite{Becchetti:2020wof}.
The results were found to be compatible with the $+5.4\%$ enhancement predicted by previous calculations utilising the soft approximation~\cite{Bonetti:2017ovy} or the $m_H \ll m_V$~\cite{Anastasiou:2008tj} and $m_V \ll m_H$~\cite{Anastasiou:2018adr} limits.

In Ref.~\cite{Czakon:2021yub} the exact top-quark mass dependence was computed to \NNLOQCD.
The challenging two-loop real corrections and three-loop virtual amplitudes are computed by numerically solving a system of differential equations~\cite{Czakon:2020vql}.
The virtual corrections are also known analytically at leading colour~\cite{Prausa:2020psw}.
The inclusion of the top quark mass corrections at \NNLOQCD shifts the cross section by $-0.26\%$ and effectively eliminates the uncertainty due to the top quark mass effects.
A study of the top-quark mass renormalisation scheme uncertainty at \NNLOQCD for off-shell Higgs production was presented in Ref.~\cite{Mazzitelli:2022scc}.

The amplitude for the production of a Higgs boson in gluon fusion ($gg \rightarrow H$) via a fermion loop is suppressed by the quark mass, $m_q$, and vanishes in the limit $m_q \rightarrow 0$.
Mass/power-suppressed non-Sudakov logarithms of the form $ y_q m_q \alpha_s^n \ln^{2n-1}(\frac{m_H}{m_q})$, where $y_q$ is the Yukawa coupling, are present in this limit and can become large.
In Refs.~\cite{Liu:2017vkm,Liu:2017axv,Liu:2018czl,Anastasiou:2020vkr}, the next-to-leading power, $\mathcal{O}(m_q)$, corrections were studied and resummed to all orders in the strong coupling constant. In Ref.~\cite{Liu:2021chn}, the next-to-next-to-leading power, $\mathcal{O}(m_q^3)$, term was obtained for 3-loop Higgs production and an all-order analysis was performed for the large-$N_c$ and Abelian limits.

Finally, although not strictly a $H$ production calculation, we mention that the recently completed calculation of the 4-loop form factors~\cite{Henn:2016men, vonManteuffel:2016xki, Henn:2016wlm, Lee:2017mip, Lee:2019zop, vonManteuffel:2019wbj, vonManteuffel:2020vjv, Agarwal:2021zft, Lee:2021uqq, Lee:2022nhh} is a first step towards $gg \rightarrow H$ at N$^4$LO.
The cusp anomalous dimension, related to the $1/\epsilon^2$ poles of the form factor, was previously computed at the same perturbative order.
Following earlier numerical results~\cite{Moch:2017uml,Moch:2018wjh}, the result for the fermionic piece was obtained~\cite{Lee:2019zop,Henn:2019rmi}, followed by the $n_f T_F C_R C_F C_A$ term~\cite{Bruser:2019auj}. The gluonic piece was computed in Ref.~\cite{Henn:2019swt}.
An independent calculation of the cusp anomalous dimension, performed without relying on any conjectured properties, was presented in Ref.~\cite{vonManteuffel:2020vjv} building on Ref.~\cite{Huber:2019fxe}.
An approximate result for the 4-loop QCD corrections to Higgs boson production was presented in Ref.~\cite{Das:2020adl}, based on the soft-gluon enhanced contributions in the limit of a large number of colours.

The experimental uncertainty on the total Higgs boson cross section is currently
of the order of 8\%~\cite{ATLAS:2019mju}
based on a data sample of 139~fb$^{-1}$,
and is expected to reduce to the order of 3\% or less with a data sample
of 3000~fb$^{-1}$~\cite{Campbell:2017hsr}. Most Higgs boson couplings will be known to 2-5\%~\cite{Cepeda:2019klc}. 
To achieve the desired theoretical uncertainty, it may be necessary to also consider the
finite-mass effects at \NNLOQCD from $b$ and $c$ quarks, combined with fully differential \NNNLOHTL corrections.

\item[$H+j$:]

\textit{LH19 status}:
Known at \NNLOHTL~\cite{Chen:2014gva,Chen:2016zka,Boughezal:2015dra,Boughezal:2015aha,Caola:2015wna,Campbell:2019gmd} and at \NLOQCD including top-quark mass effects~\cite{Jones:2018hbb,Lindert:2018iug,Neumann:2018bsx};
top--bottom interference effects are also known~\cite{Melnikov:2016qoc,Lindert:2017pky}.
Fiducial cross sections for the four-lepton decay mode were calculated in Ref.~\cite{Chen:2019wxf}.
The Higgs $p_T$ spectrum with finite quark mass effects calculated beyond the LO using high-energy resummation techniques at \LL accuracy~\cite{Caola:2016upw}; parton shower predictions including finite mass effects available in various approximations~\cite{Frederix:2016cnl,Neumann:2016dny,Hamilton:2015nsa,Buschmann:2014sia}.
The transverse momentum spectrum has also been studied at \NLOgen\!+\,\NNLL in the case a jet veto, $p_t^j \ll p_t^{j,v}$, is applied~\cite{Monni:2019yyr}.
The leading EW effects for the $qg$ and $q\bar{q}$ channels were computed some time ago~\cite{Mrenna:1995cf,Keung:2009bs}.
\medskip 

Very recently, the \NLOQCD corrections to $H+j$ production were calculated including the full mass dependence in both the bottom and top quark loops~\cite{Bonciani:2022jmb}. 
The master integrals were evaluated using the series expansion of differential equations~\cite{Moriello:2019yhu,Hidding:2020ytt,Bonciani:2016qxi,Bonciani:2019jyb,Frellesvig:2019byn}.
At the NLO level, the contribution of the bottom quark is found to be small at the inclusive level, but affects the shape of $p_T$ distribution for small $p_T$.
The process was studied with the quark masses renormalised in the on-shell (OS) and $\overline{\mathrm{MS}}$ schemes, previously full NLO results had only been presented in the OS scheme.
At large-$p_T$ the $\overline{\mathrm{MS}}$ result was found to lie below the OS result both at LO and (to a lesser extent) NLO.
Thus, the choice of mass renormalsiation scheme can have a non-negligible impact on the prediction.

Considerable work has been carried out computing the mixed QCD-EW corrections to $H+j$ production. As in $H$ production, these corrections are largely dominated by topologies containing a light fermion loop with the Higgs boson coupled to EW bosons. In~\cite{Becchetti:2018xsk}, the relevant two-loop planar master integrals were computed using differential equations. In Ref.~\cite{Bonetti:2020hqh}, the two-loop mixed corrections to $gg \rightarrow Hg$ were presented, the master integrals were computed using direct integration over the Feynman parameters and, independently, using differential equations. In Ref.~\cite{Becchetti:2021axs}, the $gg \rightarrow Hg$ amplitudes were evaluated using generalised power series, providing fast and stable numerical results. Most recently, in Ref.~\cite{Bonetti:2022lrk} the two-loop amplitudes for the $qg$ and $q\bar{q}$ channels were computed again using direct integration over the Feynman parameters. 

In Ref.~\cite{Mondini:2021nck}, the process $b\bar{b} \rightarrow H + j$ was computed differentially at \NNLOQCD in the 5FS (but retaining the bottom quark Yukawa coupling). 
The LO and NLO results in the 5FS are plagued by large factorisation scale uncertainties which are tamed at \NNLOQCD.
Although the contribution of this channel is small, it is interesting due to its sensitivity to the bottom quark Yukawa coupling. 

The production of a Higgs boson in association with a charm jet was studied at \NLOQCD in Ref.~\cite{Bizon:2021nvf}, the amplitude receives contributions from diagrams with and without a $H c \bar{c}$ Yukawa coupling and their interference. A careful treatment of the charm mass is required for this process due to the assumption $m_c = 0$ for incoming quarks (from the PDF) and the requirement of a helicity flip for the interference contribution. A related subtlety was observed also in \eg $WH(\rightarrow b\bar{b})$ when neglecting the bottom quark mass~\cite{Caola:2017xuq}.

The current experimental uncertainty on the Higgs + $\ge$ 1 jet differential cross section is of the order of 10--15\%, dominated by
the statistical error, for example the fit statistical errors for the case of the combined $H\rightarrow \gamma \gamma$ and $H\rightarrow 4\ell$ analyses~\cite{ATLAS:2022fnp,ATLAS:2020wny}.
With a sample of
3000 fb$^{-1}$ of data, the statistical error will nominally decrease by about a factor of 5, resulting in a statistical error of
the order of 2.5\%.  If the remaining systematic errors
(dominated for the diphoton analysis by the spurious signal systematic error) remain the same,
the resultant systematic error would be of the order of 9\%, leading to a
total error of approximately 9.5\%.  This is similar enough to the current theoretical uncertainty that it may motivate
improvements on the $H+j$ cross section calculation. Of course, any improvements in the systematic errors would reduce the experimental uncertainty further.
Improvements in the theory could entail a combination of the \NNLOHTL results with the
full \NLOQCD results,
similar to the reweighting procedure that has been done one perturbative order lower.

\item[$H+\geq 2j$:]

\textit{LH19 status}:
VBF production known at \NNNLOHTL accuracy for the total cross section~\cite{Dreyer:2016oyx} and at \NNLOHTL accuracy differentially~\cite{Cacciari:2015jma,Cruz-Martinez:2018rod} in the ``DIS'' approximation~\cite{Han:1992hr}; non-factorizable QCD effects beyond this approximation studied in Refs.~\cite{Liu:2019tuy}.
Full \NLOQCD corrections for $H + 3j$ in the VBF channel available~\cite{Campanario:2013fsa,Campanario:2018ppz}.
$H + \le 3j$ in the gluon fusion channel was studied in Ref.~\cite{Greiner:2016awe} and an assessment of the mass dependence of the various jet multiplicities was made in Ref.~\cite{Greiner:2015jha};
\NLOEW corrections to stable Higgs boson production in VBF calculated~\cite{Ciccolini:2007jr} and available in {\sc Hawk}~\cite{Denner:2014cla}.
Mass effects in $H+2j$ at large energy are known within the ``High Energy Jets'' framework\cite{Andersen:2009nu,Andersen:2009he,Andersen:2011hs,Andersen:2017kfc,Andersen:2018tnm,Andersen:2018kjg}.
\medskip 

In Ref.~\cite{Jager:2020hkz} parton-shower and matching uncertainties for VBF Higgs production were studied in detail using PYTHIA and HERWIG.
The study found that varying just the renormalisation, factorisation and shower scales underestimates the theoretical uncertainty.
Instead, by comparing different parton shower Monte Carlos the authors observe differences at the level of $10\%$ for NLO accurate observables and $20\%$ for LO accurate observables.
The work also highlighted the importance of the choice of appropriate recoil schemes in order not to obtain unphysical enhancements for VBF topologies.

\NNLOQCD corrections to VBF Higgs production with $H\rightarrow b\bar{b}$ and $H\rightarrow WW^*$ decays were computed for fiducial cross sections in Ref.~\cite{Asteriadis:2021gpd}, using the nested soft-collinear subtraction scheme.
These results have recently been extended to include also anomalous $HVV$ interactions~\cite{Asteriadis:2022ebf}.

A comparative study of VBF Higgs production at fixed order and with parton shower Monte Carlos has been carried out over a wide range of Higgs boson transverse momenta~\cite{Buckley:2021gfw}. This was an outgrowth of Les Houches 2019. One interesting discovery is that, at very high Higgs boson $p_T$, current implementations of ME+PS Monte Carlos do not provide a completely accurate description of the VBF production mechanism. Rather than the nominal $2 \rightarrow 3$ process, high-$p_T$ VBF Higgs production becomes effectively a $2 \rightarrow 2$ process, with the second tagging jet becoming soft with respect to the hard scattering scale. This then requires the use of two factorization scales in the ME+PS VBF calculation to take into account this disparity.

The non-factorisable \NNLOQCD correction to VBF production was studied and found to be small in Ref.~\cite{Dreyer:2020urf}.

The impact of the top-quark mass in $H+1,2$ jets was studied in Ref.~\cite{Chen:2021azt}.
For $H+1$ jet, good agreement with the full \NLOQCD result was observed when including the top-quark mass in the real radiation and rescaling the virtual contribution in the HTL by the full Born result.
\NLO{} differential predictions for $H+2$ jet were computed using this approximation and the relative correction was found to be very similar to the \NLOHTL prediction, although the absolute predictions differed significantly.

The current experimental error on the $H+\geq 2j$ cross section is on the order of
25\%~\cite{ATLAS:2022fnp}, again dominated by statistical errors,
and again for the diphoton final state, by the fit statistical error. With the same assumptions
as above, for 3000 fb$^{-1}$, the statistical error will reduce to the order of 3.5\%.
If the systematic errors remain the same, at approximately 12\% (in this case the largest
systematic error is from the jet energy scale uncertainty and the jet energy resolution uncertainty),
a total uncertainty of approximately 12.5\% would result, less than the
current theoretical uncertainty.
To achieve a theoretical uncertainty less than this value would require the calculation
of $H+\geq 2j$ to \NNLOHTL$\!\otimes\,$\NLOQCD in the gluon fusion production mode.

\item[$VH$:]

\textit{LH19 status:}
Total cross section known in the threshold limit at \NNNLOQCD~\cite{Kumar:2014uwa}. Inclusive \NNLOQCD corrections available in {\sc VH@NNLO}~\cite{Brein:2003wg,Brein:2011vx,Brein:2012ne}. 
\NNLOQCD differential results known for $WH$~\cite{Ferrera:2011bk} and $ZH$~\cite{Ferrera:2014lca}; extended to include \NNLOQCD $H\to b\bb$ decays in Ref.~\cite{Ferrera:2017zex}; matched to parton shower using the MiNLO procedure in Ref.~\cite{Astill:2016hpa,Astill:2018ivh}; supplemented with \NNLLp resummation in the 0-jettiness variable and matched to a parton shower within the \Geneva Monte Carlo framework in Ref.~\cite{Alioli:2019qzz}.
\NNLOQCD with $H\to b\bar{b}$ decays at \NNLOQCD known~\cite{Caola:2017xuq} and available in MCFM~\cite{Campbell:2016jau} with \NLOQCD decays.
Soft-gluon resummation effects known~\cite{Dawson:2012gs};
\NLOEW corrections calculated~\cite{Ciccolini:2003jy,Denner:2011id,Obul:2018psx,Granata:2017iod} also including parton shower effects~\cite{Granata:2017iod}.
Loop-induced $gg\to ZH$ known at \NLOHTL reweighted by the full LO cross section~\cite{Altenkamp:2012sx}; finite $m_t$ effects at \NLOQCD known in a $1/m_t$ expansion~\cite{Hasselhuhn:2016rqt}; threshold resummation calculated in Ref.~\cite{Harlander:2014wda}.
\NLOQCD with dimension-six Standard Model Effective Field Theory (SMEFT) operators investigated~\cite{Degrande:2016dqg}, matched to a parton shower in the \MadgraphaMCatNLO framework.
Higgs pseudo-observables investigated at \NLOQCD~\cite{Greljo:2017spw}.
Process $pp \rightarrow VH + X \rightarrow l\bar{l} b\bar{b} + X$ studied at \NNLOQCD in Ref.~\cite{Gauld:2019yng}.
Process $b\bb\to ZH$ in the 5FS, but with a non-vanishing bottom-quark Yukawa coupling, investigated in the soft-virtual approximation at \NNLOQCD~\cite{Ahmed:2019udm}.
\medskip 

The $pp \rightarrow WH(\rightarrow b\bar{b})$ process has been computed at \NNLOQCD including bottom quark mass effects~\cite{Behring:2020uzq}, the inclusion of the quark masses affects fiducial cross sections at the $5\%$ level with larger differences visible in some differential distributions.
Anomalous $HVV$ couplings were studied at \NNLOQCD for $W^\pm H$ and $ZH$ in Ref.~\cite{Bizon:2021rww}.
Predictions for $ZH$ and $W^\pm H$ with $H\rightarrow b \bar{b}$ at \NNLOQCD for production and decay were produced, matched to a parton shower using the MiNNLO method, were presented in Ref.~\cite{Zanoli:2021iyp}.
In the SMEFT, a \NNLOQCD event generator for $pp \rightarrow Z(\rightarrow l\bar{l}) H(\rightarrow b\bar{b})$ was presented in Ref.~\cite{Haisch:2022nwz}.
The polarised $q\bar{q} \rightarrow ZH$ amplitudes were studied at \NNLOQCD in Ref.~\cite{Ahmed:2020kme}.

The loop-induced $gg \rightarrow ZH$ channel accounts for $\sim 10\%$ of the total cross section and contributes significantly to the $pp \rightarrow ZH$ theoretical uncertainty. 
The NLO virtual amplitudes were computed in a small-$p_T$ expansion~\cite{Alasfar:2021ppe}, high-energy expansion~\cite{Davies:2020drs}, and numerically~\cite{Chen:2020gae}. 
The complete NLO corrections were recently presented in Ref.~\cite{Wang:2021rxu} (based on a small-$m_H,m_Z$ expansion), in Ref.~\cite{Chen:2022rua} (based on a combination of the numerical results and high-energy expansion), and in Ref.~\cite{Degrassi:2022mro} (based on a combination of the small-$p_T$ and high-energy expansion~\cite{Bellafronte:2022jmo}).

Published results for the $VH$ cross section are available for data samples up to 139\,fb$^{-1}$,
with uncertainties on the order of 20\%, equally divided between statistical and
systematic errors~\cite{ATLAS:2020fcp}. For 3000\,fb$^{-1}$, the statistical error will reduce
to 4--5\%, resulting in a measurement that is systematically limited, unless there are significant improvements to the systematic errors.
The general $VH$ process has been calculated to \NNLOQCD,
leading to a small scale uncertainty. 
However, for the best description of the $ZH$ process, the exact NLO corrections to the $gg\rightarrow ZH$ sub-process, described above, should be included.

\item[$VH+j$:]

\textit{LH19 status}: Known at \NLOQCD + PS~\cite{Luisoni:2013cuh} and \NLOSM + PS~\cite{Granata:2017iod}.
\medskip 

The $VH+j$ processes are now known at \NNLOQCD differentially, including fiducial cross sections~\cite{Gauld:2020ced,Gauld:2021ule}. The \NNLOQCD Drell-Yan type corrections are found to stabilise the predictions and reduce the theoretical uncertainty for all channels ($W^+,W^-$ and $Z$).

\item[$HH$:]

\textit{LH19 status:}
\NNNLOHTL corrections are known in the infinite top mass limit~\cite{Chen:2019lzz,Banerjee:2018lfq} and have been reweighted by the \NLOQCD result (including finite top-quark mass effects)~\cite{Chen:2019fhs}.
Finite $m_t$ effects are incorporated in \NNLOHTL calculation by reweighting and combined with full-$m_t$ double-real corrections in Ref.~\cite{Grazzini:2018bsd}.
\NLOQCD results including the full top-quark mass dependence are known numerically~\cite{Borowka:2016ehy,Borowka:2016ypz,Baglio:2018lrj,Baglio:2020ini} and matched to parton showers~\cite{Heinrich:2017kxx,Jones:2017giv}; 
exact numerical results have also been supplemented by results obtained in a small-$m_t$ expansion~\cite{Davies:2019dfy,Davies:2018qvx}; a Pad{\'e} approximated result based on the large-$m_t$ expansion and analytic results near the top threshold was presented in Ref.~\cite{Grober:2017uho}.
Threshold resummation was performed at \NLOHTL\!+\,\NNLL~\cite{Shao:2013bz} and \NNLOHTL\!+\,\NNLL~\cite{deFlorian:2015moa}.
\NLOHTL\!+\,\NLL{} resummation for the $p_T$ of the Higgs boson pair was presented in~\cite{Ferrera:2016prr}.
\NNLOQCD virtual and real-virtual corrections (involving three closed top-quark loops) known in a large-$m_t$ expansion~\cite{Grigo:2015dia,Davies:2019djw}.
Sensitivity of $HH$ production to the quartic self-coupling (which enters via EW corrections) was studied in Refs.~\cite{Liu:2018peg,Bizon:2018syu,Borowka:2018pxx}.
The $b\bb\to HH$ process is known at \NNLOQCD~\cite{Ajjath:2018ifl}.
\medskip 

The uncertainty related to the choice of the top quark mass renormalisation scheme (OS vs $\overline{\mathrm{MS}}$) at \NLOQCD is large and similar in size to the usual scale uncertainties~\cite{Baglio:2018lrj,Baglio:2020ini,Baglio:2020wgt}.
This uncertainty dominates the uncertainty budget when the \NLOQCD results are used to reweight the \NNLOHTL and \NNNLOHTL results. 

Corrections have been computed at \NLO{} within a non-linear Effective Field Theory~\cite{Heinrich:2020ckp} and used to reweight the \NNLOHTL results~\cite{deFlorian:2021azd}.
SMEFT predictions and uncertainties for $gg \rightarrow HH$ were studied at \NLOQCD{} in Ref.~\cite{Heinrich:2022idm}.

At the amplitude level, \NNLOgen results for the real corrections are now known in a large-$m_t$ expansion to order $1/m_t^6$. Results at \NNLOHTL for di-Higgs and di-pseudoscalar-Higgs production through quark annihilation have been computed \cite{Ahmed:2021hrf}.

The experimental limits on $HH$ production are currently at the level of approximately four times the
SM cross section for ATLAS~\cite{ATLAS:2020fcp} (with an expected limit  of 5.7) based on a data sample of 139\,fb$^{-1}$. The observed (expected) constraints on the Higgs boson trilinear coupling modifier $\kappa_\lambda$ are determined to be $[−1.5,6.7]$ ($[−2.4,7.7]$) at 95\% confidence level, where the expected constraints on $\kappa_\lambda$ are obtained excluding $pp\to HH$ production from the background hypothesis. For CMS, a 95\% CL limit of 3.9 times the Standard Model has been obtained~\cite{CMS:2022cpr}, with an expected limit of 7.9, for a data sample of 138\,$fb^{-1}$.
Constraints have also been set on the modifiers of the Higgs field self-coupling $\kappa_\lambda$ with this measurement in the range of $-2.3$ to 9.4,  with an expected range of $-5.0$ to 12.0. 

With a data sample of 3000 fb$^{-1}$, it is projected that a limit of
$0.5 < \lambda_{hhh}/\lambda_{hhh,\mathrm{SM}} < 1.5$ can be achieved at the $68\%$ CL
for ATLAS and CMS combined~\cite{Cepeda:2019klc}.

\item[$HH + 2j$:]

\textit{LH19 status:}
\sloppy
Fully differential results for VBF $HH$ production are known at \NNLOQCDVBFstar~\cite{Dreyer:2018rfu} and at \NNNLOQCDVBFstar for the inclusive cross section~\cite{Dreyer:2018qbw}.
\medskip 

In Ref.~\cite{Dreyer:2020urf} the non-factorisable \NNLOQCD contribution to VBF $HH$ was studied in the eikonal approximation.
For typical selection cuts, the non-factorisable \NNLOQCD corrections are found to be small
and largely contained within the scale uncertainty bands of the factorizable calculation.
It is worth emphasising that non-factorisable corrections can only be trusted in the range of validity of the eikonal approximation which is when all transverse momentum scales are small with respect to the partonic centre-of-mass energy.
It implies that the approximation does not apply when the transverse momentum of a jet becomes large.

Differential predictions for VBF $HH$ production including \NNLOQCDVBFstar + \NLOEW corrections were presented in Ref.~\cite{Dreyer:2020xaj}.
The \NLOEW corrections were found to be similar in size to the \NLOQCD corrections for typical LHC fiducial cuts and they become dominant for certain kinematic regions.
The non-factorisable \NNLOQCD contributions, computed in Ref.~\cite{Dreyer:2020urf}, are also included.
The current level of theoretical precision is adequate for the HL-LHC.

\item[$HHH$:]

\textit{LH19 status:}
Known at \NNLOHTL~\cite{deFlorian:2016sit,deFlorian:2019app}, finite quark mass effects are included by reweighting with the full Born result.

\item[$t\bar{t}H$:] 

\textit{LH19 status:}
\NLOQCD corrections for on-shell $t\tb H$ production known~\cite{Beenakker:2001rj,Reina:2001sf,Beenakker:2002nc,Dawson:2003zu}.
\NLOEW~corrections studied within the \MadgraphaMCatNLO
framework~\cite{Frixione:2014qaa,Frixione:2015zaa}.
Combined \NLOQCD and \NLOEW corrections with NWA top-quark decays computed in Ref.~\cite{Zhang:2014gcy}. 
Corrections to $t\bar{t}H$ including top quark decays and full off-shell effects
computed at \NLOQCD~\cite{Denner:2015yca},
and combined with \NLOEW~\cite{Denner:2016wet}.
\NLOQCD results merged to parton showers~\cite{Garzelli:2011vp,Hartanto:2015uka} and \NLOgen\!+\,\NNLL resummation performed in
Refs.~\cite{Kulesza:2015vda,Broggio:2015lya,Broggio:2016lfj,Kulesza:2017ukk}.
\NLOQCD results in the SMEFT calculated~\cite{Maltoni:2016yxb}.
\medskip 

In Ref.~\cite{Catani:2021cbl} results for the flavour off-diagonal channels of $t\bar{t}H$ were presented at \NNLOQCD using an extension of the $q_T$-subtraction formalism to $Q\bar{Q}F$ ($F$ colourless) final states. 
The corrections we found to be at the few per mille level for the off-diagonal channels.
The complete \NNLOQCD result is currently unknown due to the missing two-loop amplitudes.

An independent computation of \NLOQCD corrections including off-shell effects was performed in Ref.~\cite{Stremmer:2021bnk}, further considering the LO decays of the Higgs boson.

Fragmentation and splitting functions for the final-state transitions $t \rightarrow H$ and $g \rightarrow H$, known previously in the $m_H \ll m_T$ limit~\cite{Dawson:1997im} and at LO~\cite{Braaten:2015ppa}, were presented at $\mathcal{O}(y_t^2 \alpha_s)$ in Ref.~\cite{Brancaccio:2021gcz}. 
These results are useful for the resummation of logarithms of the form $\ln(p_T/m)$, and can be used to derive massive predictions in the high-$p_T$ regime from their massless counterparts.

The cross section for $t\bar{t}H$ has been measured with a data sample of 139\,fb$^{-1}$,
with a total uncertainty on the order of 20\%, dominated by the  statistical  error~\cite{ATLAS:2020ior,CMS:2020mpn}.
The statistical error will shrink to the order of 4--5\% for 3000\,fb$^{-1}$,
leaving a systematics-dominated measurement. Given that this calculation is currently known
only at \NLOQCD, with a corresponding scale uncertainty of the order of 10--15\%,
this warrants a calculation of the process to \NNLOQCD.

\item[$tH$:] 

\textit{LH19 status:}
\NLOQCD corrections known~\cite{Campbell:2013yla,Demartin:2015uha}.
\medskip

NLO QCD+EW corrections are now available~\cite{Pagani:2020mov} for on-shell top quarks.
This is the first time that NLO EW corrections have been computed for this process.
In addition, a detailed comparison between the 4-flavour and 5-flavour scheme has been carried out.

\item[$b\bar{b}H$:] (including $H$ production in bottom quark fusion treated in 5FS)

\textit{LH19 status:}
\NNLOQCD predictions in the 5FS known for a long time, inclusively~\cite{Harlander:2003ai} and later differentially~\cite{Harlander:2011fx,Buhler:2012ytl}; resummed calculation at \NNLOgen\!+\,\NNLL available~\cite{Harlander:2014hya}.
Three-loop $Hb\bar{b}$ form factor known~\cite{Gehrmann:2014vha}; \NNNLOQCD in threshold approximation~\cite{Ahmed:2014cha,Ahmed:2014era} calculated. 
Complete \NNNLOQCD results in the 5FS presented in Ref.~\cite{Duhr:2019kwi} and a resummed calculation up to \NNNLOgen\!+\,\NNNLL was presented in Ref.~\cite{Ajjath:2019neu}.
\NLOmixQED{1}{1} as well as \NNLOQED predictions were derived in Ref.~\cite{Ajjath:2019ixh}.
\NLOQCD corrections in the 4FS known since long ago~\cite{Dittmaier:2003ej,Dawson:2003kb}; 
\NLOQCD (including the formally \NNLOHTL $y_t^2$ contributions) using the 4FS presented in Ref.~\cite{Deutschmann:2018avk}.
\NLOQCD matched to parton shower and compared to 5FS in Ref.~\cite{Wiesemann:2014ioa}; various methods proposed to combine 4FS and 5FS predictions~\cite{Harlander:2011aa,Bonvini:2015pxa,Forte:2015hba,Bonvini:2016fgf,Forte:2016sja}; \NLOEW corrections calculated\cite{Zhang:2017mdz}.
\medskip 

In Ref.~\cite{Pagani:2020rsg} $b\bar{b}H$ was computed at $\mathcal{O}(\alpha_s^m \alpha^{n+1})$ with $m+n=2,3$ in the 4FS (i.e. at NLO including both QCD and EW corrections). 
New corrections from $Z(\rightarrow b\bar{b})H$ and $ZZ$ VBF were found to give sizeable corrections, making the extraction of $y_b$ from this channel considerably more challenging.
In Ref.~\cite{Grojean:2020ech}, it was shown that the impact of the new channels on the extraction of $y_b$ can be reduced using kinematic shapes.

In Ref.~\cite{Badger:2021ega}, the two-loop leading colour planar helicity amplitudes for $b\bar{b}H$ production in the 5FS were computed.
The helicity amplitudes were analytically reconstructed using finite field methods and the integrals appearing are evaluated using generalised series expansions~\cite{Hidding:2020ytt}.
The massless 4-loop QCD corrections to the $b\bar{b}H$ vertex were studied in Ref.~\cite{Chakraborty:2022yan}, this result is an important step towards N$^4$LO  $b\bar{b}\rightarrow H$ production (in the 5FS) and $H \rightarrow b \bar{b}$ decay.

\end{itemize}

\subsection{Jet final states}
An overview of the status of jet final states is given in Table~\ref{tab:SM_wishlist:wljets}.

\begin{table}
  \renewcommand{\arraystretch}{1.5}
\setlength{\tabcolsep}{5pt}
  \centering
  \begin{tabular}{lll}
    \hline
    \multicolumn{1}{c}{process} & \multicolumn{1}{c}{known} & \multicolumn{1}{c}{desired} \\
    \hline
    $pp\to 2$\,jets &
    \begin{tabular}{l}
      \NNLOQCD \\
      \NLOQCD\!+\,\NLOEW
    \end{tabular} &
    \begin{tabular}{cl}
      \NNNLOQCD\!+\,\NLOEW \\
    \end{tabular} \\
    \hline
    $pp\to 3$\,jets &
    \begin{tabular}{l}
      \NNLOQCD\!+\,\NLOEW
    \end{tabular} &
    \begin{tabular}{l}
      \\
    \end{tabular} \\
    \hline
  \end{tabular}
  \caption{Precision wish list: jet final states.}
  \label{tab:SM_wishlist:wljets}
  \renewcommand{\arraystretch}{1.0}
\end{table}

\begin{itemize}[leftmargin=2cm]

\item[2j:] 
\textit{LH19 status:} Differential \NNLOQCD corrections at leading-colour calculated in the \NNLOjet framework~\cite{Currie:2016bfm}. 
Predictions using exact colour obtained with the sector-improved residue subtraction formalism~\cite{Czakon:2019tmo} confirming in the case of inclusive-jet production at $13\,\TeV$ and $R=0.7$ that the leading-colour approximation is well justified for phenomenological applications.
Complete NLO QCD+EW corrections available~\cite{Frederix:2016ost}.
\medskip 

Completion of the full-colour result using the antenna subtraction method~\cite{Chen:2022tpk} corroborating the small impact of sub-leading colour contributions in inclusive-jet observables, however, finding larger effects in di-jet production for the triple-differential $(p_{T,\mathrm{avg}},\, y^*,\, y^\mathrm{boost})$ setup of the CMS $8\,\TeV$ measurement.
\NNLOQCD corrections to bottom quark production were computed using the $q_T$-subtraction method~\cite{Catani:2020kkl} in the four-flavour scheme with massive bottoms.

Important three-loop amplitudes became available that would enter the calculation of jet production at \NNNLOQCD: four-quark scattering ($q\bar{q}\to Q\bar{Q}$)~\cite{Caola:2021rqz} and gluon scattering~\cite{Caola:2021izf}.

Inclusive jets can be measured in both ATLAS and CMS with 5\% uncertainty in the cross sections (in the
precision range), a precision that requires \NNLOQCD cross sections.  Global PDF fits require \NNLOQCD calculations of double and even triple differential observables,  requiring the use of full colour predictions. The measurements extend to jet transverse momenta of the order of 3--5\,TeV,  necessitating the precise calculation of EW corrections as well. Eventually, PDFs will be determined at the \NNNLOQCD level, requiring the use of \NNNLOQCD predictions for the input processes, including inclusive jet production, necessitating the calculation of di-jet production to this order. 

\item[$\geq$3j:] 

\textit{LH19 status:} \NLOQCD corrections for 3-jet~\cite{Nagy:2003tz}, 4-jet~\cite{Bern:2011ep,Badger:2012pf} and 5-jet~\cite{Badger:2013yda} known.
Full \NLOSM calculation for 3-jet production was performed using \Sherpa interfaced to \Recola in Ref.~\cite{Reyer:2019obz}.
\medskip 

Completion of the 3-jet calculation at \NNLOQCD~\cite{Czakon:2021mjy} at full colour with the exception of the two-loop virtual finite remainder that is only available at leading colour so far~\cite{Abreu:2021oya}. The inclusion of the NNLO corrections
significantly reduces the dependence of the predictions on the factorization and renormalization scales. 

Three-jet observables provide a better description of jet shapes, and have the potential for the determination of the strong coupling constant over an extended dynamic range. 
\end{itemize}

\subsection{Vector boson associated processes}

An overview of the status of vector-boson associated processes is given in Table~\ref{tab:SM_wishlist:wlV}.
If not stated explicitly, the leptonic decays are assumed.
In the same way, the off-shell description is the default one.
Finally, in some cases for $VV+2j$, it is mentioned to which underlying Born contribution the corrections refer to when the full NLO corrections are not known.

\begin{table}
  \renewcommand{\arraystretch}{1.5}
\setlength{\tabcolsep}{5pt}
  \centering
  \begin{tabular}{lll}
    \hline
    \multicolumn{1}{c}{process} & \multicolumn{1}{c}{known} & \multicolumn{1}{c}{desired} \\
    \hline
    $pp\to V$ &
    \begin{tabular}{l}
      \NNNLOQCD \\
      \NLOQE11 \\
      \NLOEW
    \end{tabular} &
    \begin{tabular}{l}
      \NNNLOQCD\!+\,\NLOQE11 \\
      \NLOE2
    \end{tabular} \\
    \hline
    $pp\to VV'$ &
    \begin{tabular}{l}
      \NNLOQCD\!+\,\NLOEW{ }\wleptdecays{} \\
      \!+\,\NLOQCD{ } ($gg$ channel) \wleptdecays{} \\
    \end{tabular} &
    \begin{tabular}{l}
      \NLOQCD{ } \\($gg$ channel, w/ massive loops) \\
      \NLOQE11
    \end{tabular} \\
    \hline
    $pp\to V+j$ &
    \begin{tabular}{l}
      \NNLOQCD\!+\,\NLOEW{ }\wleptdecays{} \\
    \end{tabular} &
    \begin{tabular}{l}
      hadronic decays
    \end{tabular} \\
    \hline
    $pp\to V+2j$ &
    \begin{tabular}{l}
      \NLOQCD\!+\,\NLOEW (QCD component) \\
      \NLOQCD\!+\,\NLOEW (EW component)
    \end{tabular} &
    \begin{tabular}{l}
      \NNLOQCD \wdecays{} \\
    \end{tabular}\\
    \hline
    $pp\to V+b\bar{b}$ &
    \begin{tabular}{l}
      \NLOQCD{ }\wleptdecays{} \\
    \end{tabular} &
    \begin{tabular}{l}
      \NNLOQCD \!+\,\NLOEW{ }\wdecays{} \\
    \end{tabular} \\
    \hline
    $pp\to VV'+1j$ &
    \begin{tabular}{l}
      \NLOQCD\!+\,\NLOEW{ }\wdecays{}
    \end{tabular} &
    \begin{tabular}{l}
      \NNLOQCD \\
    \end{tabular} \\
    \hline
    $pp\to VV'+2j$ &
    \begin{tabular}{l}
      \NLOQCD \wleptdecays{} (QCD component) \\
      \NLOQCD\!+\,\NLOEW{ }\wleptdecays{} (EW component)
    \end{tabular} &
    \begin{tabular}{l}
      Full \NLOQCD\!+\,\NLOEW{ }\wdecays{} \\
    \end{tabular} \\
    \cline{2-2}
    $pp\to W^+W^++2j$ &
    \begin{tabular}{l}
      Full \NLOQCD\!+\,\NLOEW{ }\wleptdecays{} \\
    \end{tabular} &
    \begin{tabular}{l}
      \\
    \end{tabular} \\
    \cline{2-2}
    $pp\to W^+W^-+2j$ &
    \begin{tabular}{l}
      \NLOQCD\!+\,\NLOEW{ }\wleptdecays{} (EW component)\\
    \end{tabular} &
    \begin{tabular}{l}
      \\
    \end{tabular} \\
    \cline{2-2}
    $pp\to W^+Z+2j$ &
    \begin{tabular}{l}
      \NLOQCD\!+\,\NLOEW{ }\wleptdecays{} (EW component)\\
    \end{tabular} &
    \begin{tabular}{l}
      \\
    \end{tabular} \\
    \cline{2-2}
    $pp\to ZZ+2j$ &
    \begin{tabular}{l}
      Full \NLOQCD\!+\,\NLOEW{ }\wleptdecays{} \\
    \end{tabular} &
    \begin{tabular}{l}
      \\
    \end{tabular} \\
    \hline
   $pp\to VV'V''$ &
    \begin{tabular}{l}
      \NLOQCD \\
      \NLOEW{ }\wodecays{}
    \end{tabular} &
    \begin{tabular}{l}
      \NLOQCD\!+\,\NLOEW \wdecays{} \\
    \end{tabular} \\
    \cline{2-2}
   $pp\to W^\pm W^+W^-$ &
    \begin{tabular}{l}
      \NLOQCD + \NLOEW{ }\wdecays{}
    \end{tabular} &
    \begin{tabular}{l}
      \\
    \end{tabular} \\
    \hline
    $pp\to \gamma\gamma$ &
    \begin{tabular}{l}
      \NNLOQCD\!+\,\NLOEW
    \end{tabular} &
    \begin{tabular}{l}
      \NNNLOQCD \\
    \end{tabular} \\
    \hline
    $pp\to \gamma+j$ &
    \begin{tabular}{l}
      \NNLOQCD\!+\,\NLOEW
    \end{tabular} &
    \begin{tabular}{l}
      \NNNLOQCD \\
    \end{tabular} \\
    \hline
    $pp\to \gamma\gamma+j$ &
    \begin{tabular}{l}
      \NNLOQCD\!+\,\NLOEW \\
      \!+\,\NLOQCD{ }($gg$ channel)
    \end{tabular} &
    \begin{tabular}{cl}
      
    \end{tabular} \\
    \hline
    $pp\to \gamma\gamma\gamma$ &
    \begin{tabular}{l}
      \NNLOQCD
    \end{tabular} &
    \begin{tabular}{cl}
      \NNLOQCD\!+\,\NLOEW \\
    \end{tabular} \\
    \hline
  \end{tabular}
  \caption{Precision wish list: vector boson final
    states. $V=W,Z$ and $V',V''=W,Z,\gamma$.
    Full leptonic decays are understood if not stated otherwise.}
  \label{tab:SM_wishlist:wlV}
  \renewcommand{\arraystretch}{1.0}
\end{table}

\begin{itemize}[leftmargin=2cm]
\item[$V$:] 

\textit{LH19 status:}
Fixed-order \NNLOQCD and \NLOEW corrections to the Drell--Yan process known for many years, see \eg~Ref.~\cite{Alioli:2016fum} and references therein;
inclusive cross sections and rapidity distributions in the
threshold limit at \NNNLOQCD extracted from the $pp\to H$ results at this
order~\cite{Ahmed:2014cla,Ahmed:2014uya};
dominant factorizable corrections at $\mathcal{O}(\alpha_s \alpha)$ (\NLOQE11) known
differentially~\cite{Dittmaier:2015rxo} for the off-shell process including the leptonic decay;
total cross section for the $q\bar{q}$ channel at \NLOQE11 computed for on-shell Z bosons~\cite{Bonciani:2019nuy};
\NLOmixQED11 corrections for the on-shell Z boson for the inclusive cross section~\cite{deFlorian:2018wcj}, and differentially~\cite{Delto:2019ewv}.
\NNLOQCD computations matched to parton shower available using the
MiNLO method~\cite{Karlberg:2014qua}, SCET resummation \cite{Alioli:2015toa}, the UN${}^2$LOPS technique \cite{Hoche:2014uhw}, and the MINNLO${}_\text{PS}$ method~\cite{Monni:2019whf};
\NNLOQCD\!+\,\NNNLL accuracy for $\phi^\ast$ and transverse-momentum distributions of the Z boson~\cite{Bizon:2018foh};
\NNNLOQCD corrections known~\cite{Duhr:2020seh} for the production of a lepton-pair via virtual photons.
\medskip 

Several new computations became available at \NNNLOQCD accuracy:
For the neutral-current process, the lepton-pair rapidity distribution in the photon-mediated Drell-Yan process has been obtained in Ref.~\cite{Chen:2021vtu}.
The complete corrections at \NNNLOQCD to the inclusive neutral-current Drell-Yan process including contributions from both photon and Z-boson exchange has been computed in Ref.~\cite{Duhr:2021vwj}.
A computation combining a $q_T$ resummation at \NNNLL with \NNNLOQCD corrections has been presented~\cite{Camarda:2021ict}.
The same accuracy is achieved in Ref.~\cite{Chen:2022cgv}, where a detailed study of the impact of fiducial cuts on infrared physics is provided together with a critical reassessment of systematic uncertainties of such calculations.
It is worth mentioning that at this accuracy, the theoretical uncertainties are reduced to the level of $1\%$.
For the charged-current process, the total cross section at \NNNLOQCD for the charged-current Drell-Yan process were presented in Ref.~\cite{Duhr:2020sdp}, exhibiting a very similar pattern of corrections as in the case of the neutral-current process.
Most recently, differential \NNNLOQCD corrections were obtained in Ref.~\cite{Chen:2022lwc} presenting the rapidity distribution and charge asymmetry in W boson production as well as the transverse mass distribution of the decay leptons.
The latter is an important distribution for the W-boson mass extraction and \NNNLOQCD corrections were found to only minimally impact its shape.

In the last two years, there has been an extraordinary amount of work done for the computation of mixed strong-EW corrections.
In Ref~\cite{Cieri:2020ikq}, the $q_T$ formalism has been extended to the case of mixed QCD-QED corrections. The work provides the subtraction term and the hard factor needed to carry out such a computation.
Mixed QCD-electroweak corrections to on-shell Z production have been also computed by two different groups~\cite{Buccioni:2020cfi,Bonciani:2020tvf}. This work showed that QED and EW corrections displays a subtle interplay making their joint computation necessary for precision studies. 
The same accuracy has been obtained for on-shell production in the same way~\cite{Behring:2020cqi}.
The class of corrections proportional to $N_f$ have been obtained for the off-shell production W and Z~\cite{Dittmaier:2020vra}.
The corrections impact the invariant-mass distributions at a level of up to $2\%$ for large invariant masses (above $500$\,GeV).
In Ref.~\cite{Behring:2021adr}, the impact of these mixed corrections on the W-mass determination have been studied. They found that the inclusion of such corrections can impact it at the level of the order of $20$\,MeV.
An almost complete computation for the off-shell production of $pp\to\ell\nu_\ell$ have been presented in Ref.~\cite{Buonocore:2021rxx}. The only missing piece is the finite remainder of the two-loop corrections.
For the neutral Drell-Yan, such a complete computation has been obtained~\cite{Bonciani:2021zzf}.
With respect to the \NLOQCD corrections, the mixed one can reach about $−15\%$ at transverse momentum around $500$\,GeV.
In Ref.~\cite{Heller:2020owb}, an independent calculation of the two-loop EW-QCD amplitude from the one of Ref.~\cite{Bonciani:2021zzf} for the $q\bar q$-initiated Drell-Yan process has been provided.
These mixed QCD-electroweak corrections have been further studied in detail at high invariant mass in Ref.~\cite{Buccioni:2022kgy}.
Around $200$\,GeV, they are at the level of $1\%$ but can reach $3\%$ at 1\,TeV where they can be well approximated by the product of QCD and electroweak corrections.

In Ref.~\cite{Alioli:2021qbf}, a \NNNLL resummation via the RadISH formalism has been matched to \NNLOQCD corrections. This allows to generate event with parton-shower effects and hadronisation with NNLO QCD accuracy.

In Ref.~\cite{Gauld:2021pkr}, a study of transverse momentum distributions in low-mass Drell-Yan lepton pair production has been carried out at \NNLOQCD.
These have been compared to collider and fixed-target experiment data.
Only the former is described well by the calculation, indicating the importance of non-perturbative correction for the latter.

In Ref.~\cite{Frederix:2020nyw}, \NLOEW corrections to the angular coefficients parametrising the Drell-Yan process around the Z-boson pole mass.
These are provided as a function of the transverse momentum of the Z boson.

Along the same line, in Ref.~\cite{Pellen:2022fom} the corresponding decay coefficient have been provided at \NNNLOQCD+\NLOEW accuracy for the case of W production.

Drell-Yan cross sections, and in particular the production of $W$ and $Z$ bosons, are among the most precise measurements conducted at the LHC, and will continue to be so in the future. As a result, they play an important role in parton distribution function fits.  The systematic uncertainties are dominated by that of the luminosity uncertainty, with other systematic uncertainties at the percent level or smaller. The relative precision between the measured $W$ and $Z$ boson cross sections achieved at 7\,TeV by ATLAS~\cite{ATLAS:2016nqi}  has resulted in an increase of the strange quark distribution in PDF fits using that data set. Electroweak corrections are important in order to match the experimental precision, including mixed QCD and EW corrections. The data will be an important component in future PDF fits at \NNLOQCD.
The dominant luminosity uncertainty can be eliminated by considering normalised distributions, in which case an experimental uncertainty well below the percent level can be achieved in the $p_T$ distribution of the Z boson.

\item[$V/\gamma+j$:] 

\textit{LH19 status:}
$Z+j$~\cite{Gehrmann-DeRidder:2015wbt,Boughezal:2015ded,Boughezal:2016isb,Boughezal:2016yfp,Gehrmann-DeRidder:2017mvr},
$W+j$~\cite{Boughezal:2015dva,Boughezal:2016dtm,Boughezal:2016yfp,Gehrmann-DeRidder:2017mvr}, and $\gamma+j$~\cite{Campbell:2016lzl,Chen:2019zmr}
completed through \NNLOQCD including leptonic decays,
via antenna subtraction and $N$-jettiness slicing;
all processes of this class, and in particular their ratios, investigated in great
detail in Ref.~\cite{Lindert:2017olm}, combining \NNLOQCD predictions with full NLO EW and
leading \NNLOEW effects in the Sudakov approximation, including also approximations for leading
\NLOQE11 effects, devoting particular attention to error estimates and
correlations between the processes.
\medskip 

Computations for $V$ production in association with flavour jets have been obtained at \NNLOQCD: $Z+b$ \cite{Gauld:2020deh} and $W+c$ \cite{Czakon:2020coa}. One important aspect is the necessary use of the flavour-$k_{\rm T}$ algorithm in these computations in order to guarantee IR finiteness.
In Ref.~\cite{Bevilacqua:2021ovq}, a $W+c$ computation at \NLOQCD${}_\text{+PS}$ accuracy with massive charm quarks has been presented.

Polarised prediction for $W+j$ with \NNLOQCD corrections have also been obtained in Ref.~\cite{Pellen:2021vpi}. The study provides all theoretical ingredients to extract polarisation fractions and show that higher-order corrections help in their precise determination.

There are a number of kinematic variables related to $V$+jet production that probe the QCD dynamics of the hard scatter, most simply the transverse momenta of the boson and of the lead jet. At 13\,TeV, the boson and jet transverse momenta have been measured up to the order of  2\,TeV~\cite{ATLAS:2022nrp,CMS:2022ilp}. Better agreement with the data is obtained at NNLO than at NLO. Electroweak corrections are especially important for the case of the $V$ $p_T$. The $V$ $p_T$ cross section, in particular, can be measured very precisely, to the
order of a few percent. 

Note that in Ref.~\cite{Tricoli:2020uxr}, a review of recent experimental and theoretical progress have been presented for vector-boson production in association with jet(s).

\item[$V+\geq2j$:] 

\textit{LH19 status:}
\NLOQCD computations known for $V+2j$ final states
in QCD~\cite{Campbell:2002tg,Campbell:2003hd}
and EW~\cite{Oleari:2003tc}
production modes,
for $V+3j$~\cite{Ellis:2009zw,Berger:2009zg,Ellis:2009zyy,Berger:2009ep,Melnikov:2009wh,Berger:2010vm},
for $V+4j$~\cite{Berger:2010zx,Ita:2011wn} and for $W+5j$~\cite{Bern:2013gka};
\NLOEW corrections known~\cite{Denner:2014ina} including merging and showering~\cite{Kallweit:2014xda,Kallweit:2015dum};
Multi-jet merged prediction up to 9 jets at LO~\cite{Hoche:2019flt};
First results for two-loop amplitudes for $W+2j$ \cite{Hartanto:2019uvl}.
\medskip 

The leading-colour two-loop QCD corrections for the scattering of four partons and a W boson, including its leptonic decay have been computed in Ref.~\cite{Abreu:2021asb}.
The results are expressed in terms of pentagon functions, which opens the possibility to  evaluate them for the computation of \NNLOQCD cross sections.

In Ref.~\cite{Lindert:2022ejn}, \NLOQCD+\NLOEW corrections have been computed for both the $W+2j$ and $Z+2j$ processes.
The results focus on vector-boson fusion phase spaces and particular attention is devoted to the correlation of higher-order corrections in both channels.
It is particularly relevant for searches looking for invisible Higgs boson decays.

Final states with $V+\geq2j$ offer the possibility of measuring the electroweak production of a vector boson, a good testing ground for the similar formalism involved in  VBF production of a Higgs boson, as well as serving as a testbed in searches for new physics. 

\item[$V+b\bb$:] 

\textit{LH19 status:} Known at \NLOQCD for a long time~\cite{FebresCordero:2006nvf,Campbell:2008hh,FebresCordero:2009xzo,Badger:2010mg}, and matched
to parton showers~\cite{Frederix:2011qg,Oleari:2011ey,Krauss:2016orf,Bagnaschi:2018dnh};
\NLOQCD for $Wb\bb j$ calculated with parton shower matching~\cite{Luisoni:2015mpa};
$Wb\bb$ with up to three jets computed at \NLOQCD in Ref.~\cite{Anger:2017glm};
Multi-jet merged simulation, combining five- and four-flavour calculations for $Z+b\bb$ production at the LHC~\cite{Hoche:2019ncc}.
\medskip 

In Ref.~\cite{Badger:2021nhg}, an analytic computation of the two-loop QCD corrections to $ud\to W+b\bb$ for an on-shell W-boson using the leading colour and massless bottom quark approximations was presented.
This result paved the way to the first computation of $W+b\bb$ at \NNLOQCD accuracy \cite{Hartanto:2022qhh}.
This constitutes the first $2\to3$ calculation at \NNLOQCD accuracy with one massive particle.

\item[$VV'$:] 

\textit{LH19 status:} \NNLOQCD publicly available for all vector-boson
pair production processes with full leptonic decays, namely
$WW$~\cite{Gehrmann:2014fva,Grazzini:2016ctr},
$ZZ$~\cite{Cascioli:2014yka,Grazzini:2015hta,Heinrich:2017bvg,Kallweit:2018nyv},
$WZ$~\cite{Grazzini:2016swo,Grazzini:2017ckn},
$Z\gamma$~\cite{Grazzini:2013bna,Grazzini:2015nwa,Campbell:2017aul},
$W\gamma$~\cite{Grazzini:2015nwa};
\NLOQCD corrections to the loop-induced $gg$ channels
computed for $ZZ$~\cite{Caola:2015psa,Grazzini:2018owa} and
$WW$~\cite{Caola:2015rqy,Grazzini:2020stb} involving full off-shell leptonic dacays;
interference effects with off-shell Higgs contributions known~\cite{Caola:2016trd,Campbell:2016ivq};
NLO EW corrections known for
all vector-boson pair production processes including full leptonic
decays~\cite{Denner:2014bna,Denner:2015fca,Biedermann:2016yvs,Biedermann:2016guo, Biedermann:2016lvg, Biedermann:2017oae,Kallweit:2017khh,Chiesa:2018lcs}, extensively validated between several automated tools in Ref.~\cite{Proceedings:2018jsb};
combination of \NNLOQCD and \NLOEW corrections to all massive diboson processes known~\cite{Grazzini:2019jkl};
\NNLOQCD corrections to off-shell $WW$ production matched to a parton shower~\cite{Re:2018vac}.
\medskip 

In Ref.~\cite{Lombardi:2020wju}, new results for \NNLOQCD matching to parton shower applied to $Z\gamma$ have been presented. These have been further applied to a study of anomalous coupling and the background of Dark Matter searches~\cite{Lombardi:2021wug}.
The same method to match \NNLOQCD and parton-shower corrections (MiNNLOPS) has been applied to $ZZ$~\cite{Buonocore:2021fnj} and $WW$~\cite{Lombardi:2021rvg} production, respectively.

In Ref.~\cite{Cridge:2021hfr}, \NNLOQCD corrections matched to parton shower have been computed for the $W\gamma$ process.
The computation was validated against an independent calculation at fixed order and has been compared to ATLAS data at 7\,TeV, finding good agreement.

New results for \NNLOQCD predictions matched to parton shower have been presented in Ref.~\cite{Alioli:2021egp} for $ZZ$ production.
The results are based on the resummed beam-thrust spectrum and have been compared to $13$\,TeV data. Good agreement has been found with experimental data.

For $WW$ production, another formalism has been presented which resums the transverse momentum spectrum of the $WW$ pair at \NNNLL accuracy and matches it to the integrated NNLO cross section~\cite{Kallweit:2020gva}. This work highlights the importance of higher-order corrections of all type for precision phenomenology at the LHC.

In Ref.~\cite{Alioli:2021wpn}, \NLOQCD corrections matched to parton showers for the gluon--gluon loop-induced channel have been presented in the 4 leptons final state. This has been complemented by a similar study from another group where off-shell Higgs effects have been studied~\cite{Grazzini:2021iae}.

In Ref.~\cite{Chiesa:2020ttl}, exact \NLOQCD+\NLOEW corrections matched to parton shower have been presented for all massive gauge-boson channels.
Comparable results have been presented in Refs.~\cite{Brauer:2020kfv,Bothmann:2021led} were EW corrections are included approximately but the computations rely on matching and merging of higher jet multiplicities.

Several polarised predictions have been presented at \NLOQCD for $WW$~\cite{Denner:2020bcz} and $WZ$~\cite{Denner:2020eck}, at \NLOQCD+\NLOEW for $ZZ$~\cite{Denner:2021csi} and $WZ$~\cite{Le:2022lrp} production and at \NNLOQCD for $WW$~\cite{Poncelet:2021jmj}.
The corrections were found to differ for various polarisations.

Two-loop helicity amplitudes for the gluon-induced process $gg\to ZZ$ were obtained in Ref.~\cite{Agarwal:2020dye}.

As an example, it is illustrating to look at a recent measurement from the CMS collaboration~\cite{CMS:2020gtj}.
The final results states a cross section of: $\sigma_\text{tot}(pp \to ZZ) = 17.4 \pm 0.3 (\text{stat}) \pm 0.5 (\text{syst}) \pm 0.4 (\text{theo}) \pm 0.3 (\text{lumi}) \text{pb}$.
During the high-luminosity phase of the LHC, the statistical uncertainty will diminish dramatically and it is expected that systematic uncertainty will also shrink.
It will therefore leave the theory uncertainty as the dominant one.
The first source of theory uncertainty originates from the strong coupling and the PDFS, as for many other processes at the LHC.
The other theoretical source of uncertainty is the use of NLO QCD+PS tools rescaled with NNLO-QCD corrections.
The use of NNLO QCD + PS predictions, combined with EW predictions are therefore crucial for future data-theory comparisons.
In addition, given the importance of EW corrections in tails of distributions~\cite{Buonocore:2021fnj}, mixed QCD-EW corrections are likely to become relevant for data description in the future.

\item[$VV'+j$:] 

\textit{LH19 status:}
\NLOQCD corrections known for many years~\cite{Dittmaier:2007th,Campbell:2007ev,NLOMultilegWorkingGroup:2008bxd,Dittmaier:2009un,Binoth:2009wk,SM:2010nsa,Campanario:2010hp,Campanario:2009um,Campbell:2012ft,Campbell:2015hya};
\NLOEW corrections available for some on-shell processes, with subsequent leptonic decays treated
in NWA~\cite{Li:2015ura,Wang:2016pry}; full \NLOEW corrections
including decays in reach of the automated tools.
\medskip 

Fixed-order and merged parton-shower including \NLOQCD and \NLOEW corrections have been obtained for $WWj$~\cite{Brauer:2020kfv} and $ZZj$~\cite{Bothmann:2021led}. Both computations include EW corrections in an approximate way when adding PS corrections.
In Ref.~\cite{Badger:2022ncb}, the two-loop leading colour QCD helicity amplitudes for $W\gamma j$ production have been presented.

On the experimental side, measurements of this process have been performed by both experimental collaborations for example for $WWj$~\cite{ATLAS:2016agv,ATLAS:2021jgw,CMS:2020mxy}.
The experimental errors are at the level of $10\%$ or below and are dominated by systematic uncertainty.
While in the future, the statistical will eventually be negligible, the systematic errors are also expected to decrease, making therefore the total experimental error of the same order or smaller than the current theory uncertainty.
It therefore calls to go beyond current state of the art to match the accuracy of future high-luminosity measurements.

\item[$VV'+\geq2j$:]

\textit{LH19 status:} In the vector-boson scattering (VBS) approximation~\cite{Ballestrero:2018anz}, \NLOQCD corrections known for the EW production for all leptonic signatures~\cite{Jager:2006zc,Jager:2006cp,Bozzi:2007ur,Jager:2009xx,Denner:2012dz,Campanario:2013eta,Campanario:2017ffz};
Same holds true for the QCD production modes~\cite{Melia:2010bm,Melia:2011dw,Greiner:2012im,Campanario:2013qba,Campanario:2013gea,Campanario:2014ioa,Campanario:2014dpa,Campanario:2014wga};
All above computations matched to parton shower~\cite{Arnold:2008rz,Baglio:2011juf,Melia:2011gk,Jager:2011ms,Jager:2013iza,Jager:2013mu,Baglio:2014uba,Rauch:2016upa};
full \NLOSM corrections (\NLOQCD, \NLOEW and mixed \NLOgen) available for $W^+W^++2j$ production with leptonic decays~\cite{Biedermann:2017bss};
\NLOQCD+\NLOEW known for $WZ$ scattering~\cite{Denner:2019tmn};
Large EW corrections are an intrinsic feature of VBS at the LHC \cite{Biedermann:2016yds};
\NLOEW to same-sign $WW$ matched to parton/photon shower~\cite{Chiesa:2019ulk};
assessment of various approximations in VBS and parton-shower matching for same-sign WW~\cite{Ballestrero:2018anz};
\NLOQCD calculated for $WW+3j$~\cite{FebresCordero:2015kfc}.
\medskip 

In Ref.~\cite{Covarelli:2021gyz}, an extensive review of both experimental and theoretical advances in VBS has been presented.
Parton-shower effects in the EW production of $WZjj$ at \NLOQCD have been studied in Ref.~\cite{Jager:2018cyo}.
This study showcased the importance of parton-shower characteristics for jet-veto observables.
Full \NLOSM corrections became available for $ZZ+2j$ production with leptonic decays \cite{Denner:2020zit,Denner:2021hsa}.
The work showed the importance of defining fiducial phase-spaces that are stable under perturbative corrections. 
It also confirmed that large EW corrections are present for any VBS signature.
\NLOQCD+\NLOEW became available for $W^+W^-$ scattering~\cite{Denner:2022pwc}.
Interestingly, it showed that the presence of a resonant $s$-channel Higgs boson in the phase space reduces the size of the EW corrections.
Finally, Ref.~\cite{Li:2020nmi} provided an improved description of loop-induced contributions in $ZZjj$ by matching and merging different multiplicities to parton shower.

There have been several studies of the experimental prospects for VBS at the high-luminosity phase of the LHC.
They all predict that with $3000{\rm fb}^{-1}$ of data, it will be possible to measure VBS signature with a total uncertainty of a few percent \cite{CMS:2016rcn,CMS:2018zxa,ATLAS:2018uld,CMS:2021uvc}.
On the other hand, it is expected that longitudinal polarisations in such processes can only be extracted with a significance of few sigma~\cite{ATLAS:2018uld,CMS:2021uvc}.

\item[$VV'V''$:] 

\textit{LH19 status:}
\NLOQCD corrections known for many years~\cite{Hankele:2007sb,Binoth:2008kt,Campanario:2008yg,Bozzi:2009ig,Bozzi:2010sj,Bozzi:2011wwa,Bozzi:2011en,Campbell:2012ft},
also in case of $W\gamma\gamma j$~\cite{Campanario:2011ud};
\NLOEW corrections with full off-shell effects for $WWW$ production with leptonic decays \cite{Schonherr:2018jva,Dittmaier:2019twg};
\NLOEW corrections available for the
on-shell processes involving
three~\cite{Nhung:2013jta,Shen:2015cwj,Wang:2016fvj}
and two~\cite{Wang:2017wsy} massive vector bosons; $V\gamma\gamma$ processes with full leptonic decays calculated
at \NLOQCD and \NLOEW accuracy~\cite{Greiner:2017mft}.
\medskip 

\NLOEW corrections to $W\gamma\gamma$ production \cite{Zhu:2020ous} and to the production of a photon with three charged lepton plus missing energy \cite{Cheng:2021gbx} were obtained.
\NLOEW corrections to $W\gamma\gamma$ production \cite{Zhu:2020ous} and to the production of a photon with three charged lepton plus missing energy \cite{Cheng:2021gbx} were obtained.

At the moment, the measurements of the triple-production of massive bosons is statistically limited~\cite{ATLAS:2016jeu,CMS:2020hjs,ATLAS:2022xnu}.
For example, the WWW inclusive production has a $12\%$ statistical uncertainty and a $10\%$ systematic uncertainty.
Having in mind the high-luminosity phase of the LHC, at least \NLOQCD+\NLOEW accuracy will be needed to match the experimental precision.

\item[$\gamma\gamma$:] 

\textit{LH19 status:}
\NNLOQCD results for $\gamma\gamma$ production calculated by using
$q_T$ subtraction~\cite{Cieri:2015rqa,Catani:2018krb}, and by using
$N$-jettiness subtraction in the MCFM framework~\cite{Campbell:2016yrh};
\NNLOQCD also available within the public \Matrix program~\cite{Grazzini:2017mhc};
\NLOQCD corrections including top-quark mass effects to loop-induced $gg$ channel known \cite{Maltoni:2018zvp,Chen:2019fla};
$q_T$ resummation computed at \NNLL~\cite{Cieri:2015rqa};
\NLOEW corrections available for $\gamma\gamma$~\cite{Bierweiler:2013dja,Chiesa:2017gqx}.
\medskip 

Reference \cite{Gehrmann:2020oec} has presented a new \NNLOQCD calculation, relying on the antenna subtraction scheme. The study has shown that the choice of photon-isolation prescriptions as well as the choice of renormalisation and factorisation scales can have a significant impact on the predictions.
These sources of theoretical uncertainties can be relevant for comparison with experimental data.
In Ref.~\cite{Alioli:2020qrd}, \NNLL+\NNLOQCD accuracy has been achieved within the {\sc Geneva} framework.
In this way, events with parton-shower and hadronisation effects can be produced. These events are \NNLOQCD accurate for observables that are inclusive over the additional radiation.
Alternatively, in Ref.~\cite{Gavardi:2022ixt} \NNLOQCD+PS accuracy has been achieved with the help of the MiNNLOPS method.

In addition, in Ref.~\cite{Caola:2020dfu} three-loop amplitudes in QCD for di-photon production in the quark-antiquark channel have been calculated.
In particular, the helicity amplitudes obtained feature for the first time three-loop four-point function in full QCD.
This work has been completed by the tree-loop helicity amplitudes in the gluon-gluon channel~\cite{Bargiela:2021wuy}.
They provided the last missing piece for the \NNNLOQCD computation of $\gamma\gamma$ in the gluon-gluon channel.

\item[$\gamma\gamma+\ge1j$:]

\textit{LH19 status:}
\NLOQCD corrections calculated long ago~\cite{DelDuca:2003uz,Gehrmann:2013aga}, later also for
$\gamma\gamma+2j$~\cite{Gehrmann:2013bga,Badger:2013ava,Bern:2014vza} and
$\gamma\gamma+3j$~\cite{Badger:2013ava};
photon isolation effects studied at \NLOQCD~\cite{Gehrmann:2013aga};
\NLOQCD corrections for the EW production of $\gamma\gamma+2j$ \cite{Campanario:2020xaf};
\NLOEW corrections available for $\gamma\gamma j(j)$~\cite{Chiesa:2017gqx};
\medskip 

In Ref.~\cite{Chawdhry:2021hkp}, the first \NNLOQCD computation of $\gamma\gamma+j$ production has been achieved.
The calculation is exact except for the two-loop part which is computed at leading-colour accuracy.
The leading colour and light fermionic planar two-loop corrections~\cite{Agarwal:2021grm} have been computed for the $q\bar q$ and $qg$ channels by another group.
Later, the complete two-loop corrections in massless QCD for the production of two photons and a jet became available~\cite{Agarwal:2021vdh}.
In Ref.~\cite{Badger:2021ohm}, \NLOQCD to the gluon-fusion subprocess of diphoton-plus-jet production have been calculated by making use of the two-loop amplitude derived in Ref.~\cite{Badger:2021imn}.
The corrections are found to be large, justifying their inclusion when computing \NNLOQCD corrections to the quark-induced process.

\item[$\gamma\gamma\gamma$:] 

\textit{LH19 status:}
\NNLOQCD corrections in the leading-colour approximation known \cite{Chawdhry:2019bji}.
\medskip 

In Ref.~\cite{Kallweit:2020gcp}, a second computation at \NNLOQCD accuracy relying on the same approximation but employing a different subtraction method has been presented.
Both calculations are in mutual agreement and highlight that \NNLOQCD corrections are indispensable to describe experimental data.

\end{itemize}

\subsection{Top quark associated processes}

An overview of the status of top quark associated processes is given in Table~\ref{tab:SM_wishlist:wlTJ}.
\begin{table}
  \renewcommand{\arraystretch}{1.5}
\setlength{\tabcolsep}{5pt}
  \centering
  \begin{tabular}{lll}
    \hline
    \multicolumn{1}{c}{process} & \multicolumn{1}{c}{known} &
    \multicolumn{1}{c}{desired} \\
    \hline
    $pp\to t\tb$ &
    \begin{tabular}{l}
      \NNLOQCD\!+\,\NLOEW (w/o decays) \\
      \NLOQCD\!+\,\NLOEW{ }(off-shell effects) \\
      \NNLOQCD{ }(w/ decays)
    \end{tabular} &
    \begin{tabular}{l}
      \NNNLOQCD
    \end{tabular} \\
    \hline
    $pp\to t\tb+j$ &
    \begin{tabular}{l}
      \NLOQCD{ }(off-shell effects) \\
      \NLOEW (w/o decays)
    \end{tabular} &
    \begin{tabular}{l}
      \NNLOQCD\!+\,\NLOEW{ }(w/ decays)
    \end{tabular} \\
    \hline
    $pp\to t\tb+2j$ &
    \begin{tabular}{l}
      \NLOQCD{ }(w/o decays)
    \end{tabular} &
    \begin{tabular}{l}
      \NLOQCD\!+\,\NLOEW{ }(w/ decays)
    \end{tabular} \\
    \hline
    $pp\to t\tb+V'$ &
    \begin{tabular}{l}
      \NLOQCD\!+\,\NLOEW{ }(w/o decays)
    \end{tabular}
    &
    \NNLOQCD\!+\,\NLOEW{ }(w/ decays)
    \\
    $pp\to t\tb+\gamma$ &
    \begin{tabular}{l}
      \hline
      \NLOQCD{ }(off-shell effects)
    \end{tabular} & \\
    $pp\to t\tb+Z$ &
    \begin{tabular}{l}
      \hline
      \NLOQCD{ }(off-shell effects)
    \end{tabular} & \\
    $pp\to t\tb+W$ &
    \begin{tabular}{l}
    \hline
    \NLOQCD\!+\,\NLOEW{ }(off-shell effects) \\
    \end{tabular} &
    \begin{tabular}{l}
      
    \end{tabular} \\
    \hline
    $pp\to t/\tb$ &
    \begin{tabular}{l}
      \NNLOQCD{*}(w/ decays) \\
      \NLOEW{ }(w/o decays)
    \end{tabular} &
    \begin{tabular}{l}
      \NNLOQCD\!+\,\NLOEW{ }(w/ decays)
    \end{tabular} \\
    \hline
    $pp\to tZj$ &
    \begin{tabular}{l}
      \NLOQCD\!+\,\NLOEW{ }(w/ decays)
    \end{tabular} &
    \begin{tabular}{l}
      \NNLOQCD\!+\,\NLOEW{ } (w/o decays)
    \end{tabular} \\
    \hline
    $pp\to t\tb t\tb$ &
    \begin{tabular}{l}
      Full \NLOQCD\!+\,\NLOEW{ }(w/o decays)
    \end{tabular} &
    \begin{tabular}{l}
      \NLOQCD\!+\,\NLOEW{ }(off-shell effects) \\
      \NNLOQCD
    \end{tabular} \\
    \hline
  \end{tabular}
  \caption{Precision wish list: top quark  final states. \NNLOQCD$^{*}$ means a
   calculation using the structure function approximation. $V'=W,Z,\gamma$.}
  \label{tab:SM_wishlist:wlTJ}
  \renewcommand{\arraystretch}{1.0}
\end{table}

\begin{itemize}[leftmargin=2cm]

\item[$t\tb$:] 

\textit{LH19 status:}
Fully differential \NNLOQCD computed for on-shell top-quark pair production~\cite{Czakon:2015owf,Czakon:2016ckf,Czakon:2016dgf,Catani:2019hip}, also available as {\tt fastNLO} tables~\cite{Czakon:2017dip};
polarised two-loop amplitudes known~\cite{Chen:2017jvi};
combination of \NNLOQCD and \NLOEW corrections performed~\cite{Czakon:2017wor};
also multi-jet merged predictions with \NLOEW corrections available~\cite{Gutschow:2018tuk};
resummation effects up to \NNLL computed~\cite{Beneke:2011mq,Cacciari:2011hy,Ferroglia:2013awa,Broggio:2014yca,Kidonakis:2015dla,Pecjak:2016nee};
\NNLOQCD\!+\,\NNLL for (boosted) top-quark pair production~\cite{Czakon:2018nun};
top quark decays known at \NNLOQCD~\cite{Gao:2012ja,Brucherseifer:2013iv};
Complete set of \NNLOQCD corrections to top-pair production and decay in the NWA for intermediate top quarks and $W$ bosons~\cite{Behring:2019iiv}; $W^+W^- b\bar{b}$ production with full off-shell effects calculated
at \NLOQCD~\cite{Denner:2010jp,Denner:2012yc,Bevilacqua:2010qb,Heinrich:2013qaa}
including leptonic $W$ decays, and in the lepton plus jets channel~\cite{Denner:2017kzu};
full \NLOEW corrections for leptonic final state available~\cite{Denner:2016jyo};
calculations with massive bottom quarks available at
\NLOQCD~\cite{Frederix:2013gra,Cascioli:2013wga};\\
\NLOQCD predictions in NWA matched to parton shower~\cite{Campbell:2014kua}, and multi-jet merged
for up to 2 jets in \Sherpa~\cite{Hoeche:2014qda} and \Herwig\,7.1~\cite{Bellm:2017idv};
$b \bb 4\ell$ at \NLOQCD matched to a parton shower in the \Powheg framework retaining all off-shell and non-resonant contributions~\cite{Jezo:2016ujg}. 

\medskip 

The first \NNLOQCD computation matched to parton shower using the MINNLO${}_\text{PS}$ method has been presented in Ref.~\cite{Mazzitelli:2020jio,Mazzitelli:2021mmm} for on-shell top production.
The decays of the top quark are described at tree level retaining spin correlation.
Phenomenological results are also produced by comparing them against experimental data.
As a by product, events with \NNLOQCD accuracy can be generated.

In Ref.~\cite{Czakon:2021ohs}, \NNLOQCD corrections to identified heavy hadron production at hadron colliders has been provided.
As an application, the authors study B-hadron production in $t\tb$.

In Ref.~\cite{Czakon:2020qbd}, an extensive study of leptonic observables in top-quark pair production and decay at \NNLOQCD accuracy has been provided. In particular, both inclusive and fiducial predictions are studied in details and compared to experimental data.

\NNLOQCD predictions with an $\overline{\textrm{MS}}$ top-quark mass have been computed~\cite{Catani:2020tko}.
This allows effects due to the running of the $\overline{\textrm{MS}}$ mass of the top quark to be studied. 

Reference~\cite{Alioli:2021ggd} provided a zero-jettiness resummation for top-quark pair production at the LHC. 
In that way, \NNLL predictions are obtained and can be compared to different resummation predictions as well as combined with \NLOQCD predictions.

New studies of \NLOQCD+\NLOEW corrections to the top quark pair production have been provided in Ref.~\cite{Frederix:2021zsh}.
The predictions are NLO accurate for the on-shell production of the top quarks while the decay is treated at tree level.
The effect of such corrections is studied for spin-correlation coefficients and various asymmetry observables.

New results of two-loop amplitude have also emerged.
In Ref.~\cite{Badger:2021owl}, the two-loop leading colour QCD helicity amplitudes for top-quark pair production in the gluon channel have been calculated.
In particular, it provides a complete set of analytic helicity amplitudes which includes contributions due to massive fermion loops.

Top-pair production has proven to be an important process for inclusion in global PDF fits, providing additional information on the gluon distribution, especially at higher $x$. Results are available in variety of final states, depending on the decays of the two $W$ bosons.  One advantage of this process is that it offers multiple observables that can be used in  PDF fits, with statistical correlations  provided by the experiments that prevent double-counting. Measurements cover a very wide kinematic range with the top-quark pair mass currently covering a range up to 4\,TeV~\cite{ATLAS:2022mlu}, which will extend to 7\,TeV at the high-luminosity LHC. Electroweak corrections become very important at higher masses. Measurements at the higher mass range also require the resolution of boosted topologies.

\item[$t\tb\,j$:] \textit{LH19 status:}
\NLOQCD corrections calculated for on-shell top quarks~\cite{Dittmaier:2007wz,Melnikov:2010iu,Melnikov:2011qx},
also matched to parton showers~\cite{Kardos:2011qa,Alioli:2011as}; full off-shell decays
included at \NLOQCD~\cite{Bevilacqua:2015qha,Bevilacqua:2016jfk};
\NLOEW corrections known~\cite{Gutschow:2018tuk} for on-shell top quarks.
\medskip 

Very recently, a phenomenological study of $t\tb\,j+X$ has been presented~\cite{Alioli:2022lqo}.
It considers $t\tb\,j+X$ as a signal and provides theoretical ingredients for the measurement of the process and in particular the extraction of the top mass thanks to it.

\item[$t\tb+\ge2j$:]\textit{LH19 status:}
\NLOQCD corrections to $t\tb jj$  known for many years~\cite{Bevilacqua:2010ve,Bevilacqua:2011aa};
$t\tb jjj$ at \NLOQCD calculated~\cite{Hoche:2016elu} using \Sherpa{}+\OpenLoops.

\item[$t\tb+b\bb$:] \textit{LH19 status:}
\NLOQCD corrections to $t\tb b\bb$ with massless bottom quarks known
for a long time~\cite{Bredenstein:2009aj,Bevilacqua:2009zn,Bredenstein:2010rs};
\NLOQCD with massive bottom quarks and matching to a parton shower investigated~\cite{Cascioli:2013era,Jezo:2018yaf};
\NLOQCD corrections for $t\bar{t}b\bar{b}$ production in association with a
light jet~\cite{Buccioni:2019plc}; All computations performed with on-shell top quarks.
\medskip 

Two independent computations \cite{Denner:2020orv,Bevilacqua:2021cit} have obtained \NLOQCD corrections for the full off-shell $2\to8$ process, retaining all non-resonant and interference effects.

\item[$t\tb t\tb$:]

This process can serve as a probe of the Yukawa coupling of the top quark to the Higgs, as well as a background to possible new physics, such as gluino pair production~\cite{Bevilacqua:2012em}. This process was not in the 2019 Les Houches wishlist, but was the last calculation to be completed in the original NLO Les Houches wishlist~\cite{SM:2012sed}. This calculation was performed in 2012~\cite{Bevilacqua:2012em} for on-shell top quarks at \NLOQCD, reducing the scale uncertainty from the order of 60\% at LO to 20-25\% at NLO. Additional NLO EW contributions were calculated in Ref.~\cite{Frederix:2017wme} considering also subleading contributions and were found to be relatively large (albeit with large cancellations). Measurements of four-top production have been carried out by both ATLAS~\cite{ATLAS:2020hpj,ATLAS:2021kqb} and CMS~\cite{CMS:2019jsc,CMS:2019rvj}, with the ATLAS measurement reaching a significance of 4.7 sigma. The uncertainties are evenly balanced between statistical and systematic sources, with each being of the same order as the current theory uncertainty (through scale variation at NLO). Clearly, both experimental sources will decrease with more data. It is worth noting that a sizeable fraction of the systematic error is related to the signal modelling, which could be reduced by improvements in the theoretical determination.  The calculation of this process to \NNLOQCD ($2\rightarrow4$ with a heavy mass scale) is not on the current horizon (but perhaps would be feasible within the lifetime of the HL-LHC). Shorter-term improvements would include NLO top-quark decays with NLO spin-correlations.
Recently, \NLOQCD corrections matched to parton shower have been implemented in the \Powheg
framework~\cite{Jezo:2021smh}. All subleading EW production modes are included at LO and the top-quark decays are modelled at LO, retaining spin-correlation effects.

\item[$t\tb V^\prime$:] 

\textit{LH19 status:}
\NLOQCD corrections to $t\tb Z$ including NWA decays considered~\cite{Rontsch:2014cca,Rontsch:2015una};
\NLOQCD for off-shell process \cite{Bevilacqua:2018woc} and \NLOEW for on-shell top quarks \cite{Duan:2016qlc};
\NLOQCD corrections to $t\tb\gamma$~\cite{Bevilacqua:2018woc} for off-shell top quarks and \NLOEW corrections to for on-shell top quarks~\cite{Duan:2016qlc};
\NLOQCD corrections to $t\tb\gamma\gamma$ production matched to parton shower~\cite{vanDeurzen:2015cga};
\NLOEW and \NLOQCD corrections to $t\tb Z/W/H$ computed
within \MadgraphaMCatNLO~\cite{Frixione:2015zaa};
dedicated studies on complete \NLOSM corrections for $t\tb W$ and $t\tb t\tb$ production~\cite{Frederix:2017wme};
resummed calculations up to \NNLL to $t\tb W$~\cite{Broggio:2016zgg,Kulesza:2018tqz}
and $t\tb Z$~\cite{Broggio:2017kzi,Kulesza:2018tqz} production;
combination of these corrections with \NLOEW~\cite{Broggio:2019ewu} for $t\tb Z/W/H$.
\medskip 

\NLOQCD corrections to the full off-shell production of $t\tb W$ in the leptonic channels have been obtained by two different groups~\cite{Bevilacqua:2020pzy,Denner:2020hgg}. Correlations and asymmetries have been further studied in Ref.~\cite{Bevilacqua:2020srb}. Combined \NLOQCD and \NLOEW corrections to this process were completed in Ref.~\cite{Denner:2021hqi}. Finally, in Ref.~\cite{Bevilacqua:2021tzp} the impact of off-shell, parton-shower, and spin-correlation effects have been studied.
In addition, improved merging \cite{Frederix:2021agh} as well as subleading EW corrections and spin correlation \cite{Frederix:2020jzp} have been studied for $t\tb W$.
\NLOQCD corrections of the full off-shell process $t\tb Z$ has been presented in Ref.~\cite{Bevilacqua:2019cvp,Bevilacqua:2022nrm}.

The full tower of NLO corrections have been obtained for $t\tb\gamma$ in Ref.~\cite{Pagani:2021iwa}.
In the same work, NLO QCD+EW corrections have been also computed for $t\tb\gamma\gamma$ and $t\gamma j$.
Particular attention has been paid to obtaining the corrections to these processes with isolated photons in an automated framework.

In Ref.~\cite{Ghezzi:2021rpc}, an implementation in \Powheg of \NLOQCD corrections matched to parton shower for on-shell top quarks has been presented for $t\tb \ell^+\ell^-$ production.
In the same way, similar predictions were obtained in Ref.~\cite{FebresCordero:2021kcc} for $t\tb W$.

In Ref.~\cite{Kulesza:2020nfh}, \NNLL+\NLOQCD corrections have been presented for cross sections and differential distributions for $t\tb W/Z/h$.

A comparison of different predictions, including NWA and off-shell calculations at \NLOQCD was carried out~\cite{Bevilacqua:2019quz}.

\item[$t$/$\tb$:] 

\textit{LH19 status:}
Fully differential \NNLOQCD corrections for the dominant $t$-channel production process completed
in the structure function approximation, for stable top quarks~\cite{Brucherseifer:2014ama} and
later including top-quark decays to \NNLOQCD accuracy in the NWA~\cite{Berger:2016oht,Berger:2017zof};
\NNLOQCD corrections for the $s$-channel and related decay,
neglecting the colour correlation between the light and heavy quark lines and applying the NWA~\cite{Liu:2018gxa}.
\NLOQCD corrections to $t$-channel electroweak $W+bj$ production available
within MG5\_aMC@NLO~\cite{Papanastasiou:2013dta,Frixione:2005vw};
\NLOQCD corrections to single-top production in the $t, s$ and $tW$ channels also
available in \Sherpa~\cite{Bothmann:2017jfv}
and in \Powheg~\cite{Alioli:2009je,Re:2010bp};
\NLOEW corrections known~\cite{Frederix:2019ubd}.
\NLOQCD for single top-quark production in association with two jets \cite{Molbitz:2019uqz}.
\NLOQCD matched to parton shower for single top-quark production in association with a jet in the \MiNLO method~\cite{Carrazza:2018mix};
Soft-gluon resummation at \NLLone for single-top production in the $t$-channel~\cite{Cao:2019uor} and the $s$-channel modes~\cite{Sun:2018byn}.
\medskip 

An independent calculation of differential $t$-channel single-top production and decay at \NNLOQCD was performed in Ref.~\cite{Campbell:2020fhf}, resolving a disagreement found between the two prior calculations.

In Ref.~\cite{Bronnum-Hansen:2021pqc}, the non-factorisable contribution to the two-loop helicity amplitude for $t$-channel single-top production were recently computed.
These can reach a few percent at high top transverse momentum.

\item[$tZj$:] 

\textit{LH19 status:} \NLOQCD corrections known for on-shell top quarks \cite{Campbell:2013yla}.
\medskip 

In Ref.~\cite{Pagani:2020mov}, \NLOQCD+\NLOEW corrections have been computed for on-shell top quarks.
While this process is interesting on its own (it has been measured by both ATLAS \cite{ATLAS:2017dsm,ATLAS:2020bhu} and CMS \cite{CMS:2017wrv,CMS:2018sgc}), it also appears as background to vector-boson scattering for the $WZ$ signature.

\end{itemize}

\subsection*{Acknowledgements}
We thank all of our colleagues who provided us with valuable input to update the wishlist. In particular, we would like to thank Melissa van~Beekveld, Gudrun Heinrich, Stefan H\"oche, Matthew Lim, Jonas Lindert, Andreas von~Manteuffel, Simone Marzani, Davide Pagani, Giovanni Pelliccioli, German Sborlini, and Malgorzata Worek for their helpful feedback.
S.P.J.\ is supported by a Royal Society University Research Fellowship (Grant URF/R1/201268).
This work is supported in part by the UK Science and Technology Facilities Council (STFC) through grant ST/T001011/1.
M.P.\ acknowledges support by the German Research Foundation (DFG) through the Research Training Group RTG2044.

\let\NLO\undefined
\let\NLL\undefined
\let\NLOH\undefined
\let\NLOQ\undefined
\let\NLOE\undefined
\let\NLOHone\undefined
\let\NLOQone\undefined
\let\NLOEone\undefined
\let\NLOQE\undefined
\let\LOQ\undefined
\let\NLOQonetb\undefined
\let\NLOQtb\undefined
\let\NLOQmtsix\undefined
\let\NLOQzzero\undefined
\let\NLOQoneVBF\undefined
\let\NLOQVBF\undefined
\let\NLOQoneDIS\undefined
\let\NLOQDIS\undefined
\let\NLOEoneVBF\undefined
\let\NLOQoneVBFstar\undefined
\let\NLOQVBFstar\undefined
\let\NLOEoneVBFstar\undefined
\let\NLOggHVtb\undefined

\let\xs\undefined
\let\tb\undefined
\let\bb\undefined
\let\qb\undefined
\let\VdkL\undefined
\let\VdkQ\undefined
\let\VdkLNWA\undefined
\let\VdkQNWA\undefined

\let\wodecay\undefined
\let\wdecay\undefined
\let\wodecays\undefined
\let\wdecays\undefined
\let\wleptdecays\undefined

\let\VdkALLNWA\undefined
\let\VdkALL\undefined
\let\tdk\undefined
\let\tdkNWA\undefined
\let\TVdkALLNWA\undefined

\let\MadgraphaMCatNLO\undefined
\let\Herwig\undefined
\let\Powheg\undefined
\let\Powhegboxres\undefined
\let\PowhegboxVtwo\undefined
\let\GoSam\undefined
\let\Recola\undefined
\let\OpenLoops\undefined
\let\MadLoop\undefined
\let\Matrix\undefined
\let\Munich\undefined
\let\Geneva\undefined
\let\Sherpa\undefined
\let\NNLOjet\undefined
\let\MiNLO\undefined
\let\NLOX\undefined


\clearpage

\bibliography{LH21.bib}


\end{document}